\documentclass[aps, prl,10pt,twocolumn,superscriptaddress,nofootinbib]{revtex4-1}
\usepackage{mathrsfs, amssymb, amsmath}  
\usepackage{epsfig, cancel}
\usepackage{latexsym}
\usepackage{caption}
\usepackage{natbib, comment}
\usepackage{booktabs}
\usepackage{url}
\usepackage{dcolumn}
\usepackage{multirow}
\usepackage{color}
\usepackage{cancel}
\usepackage{soul}
\usepackage[normalem]{ulem}
\usepackage{amsfonts,amssymb,amsmath, txfonts}
\usepackage{graphicx,epsfig}
\usepackage{psfrag}
\usepackage{hyperref}
\hypersetup{colorlinks=true}
\usepackage{mathtools}
\usepackage{enumitem}
\usepackage{float}
\usepackage[dvipsnames]{xcolor}
\usepackage{xcolor}
\hypersetup{ linktoc=all,
    colorlinks, linkcolor={blue},
    citecolor={darkred}, urlcolor={darkgreen}
}

\usepackage{mathrsfs, amssymb, amsmath, mathtools} 
\usepackage{multirow}
\usepackage{amsfonts,amssymb,amsmath, txfonts}
\usepackage{graphicx,epsfig}
\usepackage{hyperref}
\usepackage{xcolor}
\usepackage{subcaption}
\usepackage{booktabs}
\graphicspath{{Images/}}
\newlength{\DefaultTabColSep}
\definecolor{rosy}{RGB}{230,235,252}
\definecolor{myframetitle}{RGB}{90,89,170}
\definecolor{myblocktitle}{RGB}{140,185,249}
\definecolor{mytitle}{RGB}{10,80,26}

\definecolor{darkgreen}{RGB}{27,130,45}
\definecolor{darkblue}{rgb}{0,0,0.3}
\definecolor{darkred}{rgb}{0.7,0,0}

\definecolor{light gray}{RGB}{220,220,220}
\definecolor{dark purple}{RGB}{108,0,217}
\definecolor{pink}{RGB}{190,20,100}
\definecolor{orang}{RGB}{193,63,0}
\definecolor{green}{RGB}{11,98,17}
\definecolor{darkpink}{RGB}{153,0,76}
\definecolor{bluegreen}{RGB}{0,102,102}
\definecolor{greenlagan}{RGB}{0,102,0}
\definecolor{redgreen}{RGB}{102,102,0}
\definecolor{Redgreen}{RGB}{153,76,0}
\definecolor{vividviolet}{rgb}{0.62, 0.0, 1.0}
\definecolor{amaranth}{rgb}{0.9, 0.17, 0.31}
\definecolor{palatinateblue}{rgb}{0.15, 0.23, 0.89}
\definecolor{brightpink}{rgb}{1.0, 0.0, 0.5}
\definecolor{cornflowerblue}{rgb}{0.39, 0.58, 0.93}
\definecolor{deepcarminepink}{rgb}{0.94, 0.19, 0.22}
\definecolor{radicalred}{rgb}{1.0, 0.21, 0.37}

\def\H0{{\text{H}\hspace*{-2.05mm}\text{H} 0\hspace*{-1.35mm}0\ }}

\def\be{\begin{equation}}
\def\ee{\end{equation}}
\def\beq{\begin{equation}}
\def\eeq{\end{equation}}
\def\bea{\begin{eqnarray}}
\def\eea{\end{eqnarray}}
\newcommand{\dd}{\textrm{d}}

\begin{document}

\title{Crosschecking Cosmic Distances from  DESI BAO and DES SNe}

\author{Mauricio Lopez-Hernandez}
\affiliation{Departamento de F\'isica, Centro de Investigaci\'on y de Estudios Avanzados del I.P.N.
Apartado Postal 14-740, 07000, Ciudad de M\'exico, M\'exico}
\affiliation{Atlantic Technological University, Ash Lane, Sligo F91 YW50, Ireland}
\author{Eoin \'O Colg\'ain}
\affiliation{Atlantic Technological University, Ash Lane, Sligo F91 YW50, Ireland}
\author{Saeed Pourojaghi} 
\affiliation{School of Physics, Institute for Research in Fundamental Sciences (IPM), P.O.Box 19395-5531, Tehran, Iran}
\author{M. M. Sheikh-Jabbari} 
\affiliation{School of Physics, Institute for Research in Fundamental Sciences (IPM), P.O.Box 19395-5531, Tehran, Iran}

\begin{abstract}
We perform a consistency check of DESI DR2 BAO  constraints ($D_M/r_d, D_H/r_d)$ by reconstructing the same quantities from DES supernovae (SNe) in bins with the same effective redshift $z_{\textrm{eff}} \in \{ 0.510, 0.706, 0.934 \}$ and a Planck $r_d$ prior. Through mock analysis we show that $D_M(z_{\rm eff})$ and $D_{H}(z_{\rm eff})$ can be locally reconstructed model agnostically from $\Lambda$CDM and extended models, but only if one employs frequentist methods; purely Bayesian reconstructions from Markov Chain Monte Carlo (MCMC) exhibit bias. We find that the ratio of the three $D_M/r_d$ values at different $z_{\textrm{eff}}$ are consistent with a horizontal, thus confirming that the distance duality relation holds up to calibration. However, the $D_H/r_d$ ratio shows a decreasing trend driven by the $z_{\textrm{eff}} = 0.934$ bin, the significance of which varies from $2.5 \sigma$ with Bayesian methods down to $1.4 \sigma$ with frequentist methods. We show that replacing DES with DES-Dovekie SNe reduces the significance to $1.7 \sigma$ and $1.2 \sigma$ in Bayesian and frequentist approaches, respectively. We conclude that distances reconstructed from SNe show good agreement with DESI BAO distances across the redshifts studied. We also note that $D_M(z_{\rm eff} = 0.510)/r_d$ reconstructed from SNe favours DESI BAO over transversal BAO against a backdrop of a $3.7 \sigma$ disagreement.      
\end{abstract}

\maketitle

\section{Introduction}

The Hubble constant tension has persisted for over a decade (see \cite{DiValentino:2021izs, Perivolaropoulos:2021jda, Abdalla:2022yfr, CosmoVerseNetwork:2025alb} for reviews). If systematics are not at play, the key implication is that physics is missing from the $\Lambda$CDM model. Distance indicators in the local Universe show scatter in their $H_0$ determinations \cite{Riess:2021jrx, Freedman:2021ahq, Pesce:2020xfe, Kourkchi:2020iyz, Blakeslee:2021rqi}, implying that observational biases/systematics are not fully under control. Good science demands crosschecks to ascertain whether different calibrators agree on the distance $D$ to the same galaxy. Noting that $v = c z = H_0 D$ in the Hubble-Lema\^itre Law, crosschecks of  distances at fixed redshift $z$ are being performed \cite{Li:2024pjo}. 

Recently, the Dark Energy Spectroscopic Instrument (DESI) collaboration claimed a dynamical dark energy (DE) signal \cite{DESI:2024mwx, DESI:2024hhd, DESI:2025zgx} (see \cite{DES:2024jxu, Rubin:2023ovl} for earlier claims), admittedly one that makes Hubble tension worse.\footnote{As pointed out in \cite{Lee:2022cyh}, DE models with $w_0:= w(z=0) > -1$ tend to exacerbate $H_0$ tension, but it is possible for $w(z) < -1$ at $z >0$ to compensate and increase $H_0$ relative to $\Lambda$CDM in a small class of datasets. However, for DESI data as $w_0$ increases, thus exhibiting more of a deviation from $\Lambda$CDM, $H_0$ decreases. See appendix  of \cite{Colgain:2025nzf} for explicit numbers.} Although the dynamical DE signal, characterised by $w_0 > -1$ in the $w_0 w_a$CDM model \cite{Chevallier:2000qy, Linder:2002et}, is evident in DESI DR1 \cite{DESI:2024mwx} and DR2 baryon acoustic oscillations (BAO) data alone \cite{DESI:2025zgx} (see also \cite{Colgain:2024mtg, Colgain:2025fct}), yet absent in DESI DR1 full-shape (FS) modelling alone \cite{DESI:2024jxi} because tomographic constraints are consistent with $\Lambda$CDM ($w_0 = -1, w_a = 0$), to get a statistically significant signal one must currently combine with Cosmic Microwave Background (CMB) \cite{Planck:2018vyg, ACT:2025fju} and Type Ia supernovae (SNe) datasets \cite{Brout:2022vxf, Rubin:2023ovl, DES:2024jxu}.\footnote{The signal is diluted in extended 12-parameter models \cite{RoyChoudhury:2024wri, RoyChoudhury:2025dhe}.} Given that BAO and SNe probe overlapping redshifts, before jumping to conclusions, especially once the theoretical difficulties are appreciated \cite{Wolf:2025acj}, it is once again good science to crosscheck cosmological distances before combining data.  

Our approach here is straightforward. Given that DESI provides $(D^{\textrm{DESI}}_M/r_d, D^{\textrm{DESI}}_H/r_d)$ constraints at effective redshift $z_{\textrm{eff}}$, we bin the DES SNe sample \cite{DES:2024jxu} so that the bins possess as close as possible to the same $z_{\textrm{eff}}$. Given Pantheon+ SNe \cite{Brout:2022vxf} are sparse at higher redshifts, making it difficult to construct bins, and Union3 \cite{Rubin:2023ovl} has not officially released full data, DES is currently the optimal dataset to perform the test. In each bin, we assume the (flat) $\Lambda$CDM model and the Planck value for the sound horizon radius $r_d$ to reconstruct $D_M^{\textrm{DES}}/r_d$ and $D_H^{\textrm{DES}}/r_d$, before constructing ratios of DES with DESI constraints $R_{D_X} = (D_X^{\textrm{DES}}/r_d)/(D_X^{\textrm{DESI}}/r_d)$, $X \in \{M, H\}$,  at $z_{\textrm{eff}}$. Consistency demands $R_{D_{X}}$ is a constant function of $z_{\textrm{eff}}$, which up to a choice of calibrating SNe absolute magnitude $M_B$ and $r_d$ should agree with unity. Any shift from $R_{D_X} = 1$ may be interpreted as a realisation of $H_0$ tension provided both ratios show the same shift. 

The fact that raw SNe data constrain the functional form of the luminosity distance $D_L(z)$, which is generically proportional to $D_M(z)$, while $D_H(z)$ is the first derivative of $D_M(z)$ (in a flat universe), allows us to entertain the idea that both $D_M(z)$ and $D_H(z)$ can be \textit{locally} reconstructed cosmological model independently at the effective redshift $z_{\rm eff}$ of the bin. To demonstrate this we inject a $w_0 w_a$CDM cosmology distinct from $\Lambda$CDM, where we show that both $\Lambda$CDM and $w_0 w_a$CDM agree on the reconstructed $D_M(z_{\rm eff})$ but models show a bias in the expected $D_H(z_{\rm eff})$ value with $\Lambda$CDM performing best. We trace this bias to a reconstruction method based on Markov Chain Monte Carlo (MCMC) and show that it can be corrected with frequentist profile likelihood methods \cite{Trotta:2017wnx, Gomez-Valent:2022hkb, Colgain:2023bge}. This implies that projection or volume effects impact reconstructions of $D_H(z_{\rm eff})$ from Bayesian posteriors.   

More concretely, we find that $R_{D_M} (z_{\rm eff})$ values are robust across Bayesian and frequentist methods, showing little sensitivity to the underlying model, in line with expectations from SNe constraining $D_L(z) \propto D_M(z)$. Moreover, $R_{D_M}(z_{\rm eff})$ constraints at $z_{\rm eff} \in \{0.510, 0.706, 0.934 \}$ are consistent with a horizontal. Calibration aside, our analysis precludes a problem with the distance duality relation $D_{L}(z) = (1+z)^2 D_{A}(z)$ (see \cite{Favale:2024sdq, Teixeira:2025czm, Afroz:2025iwo, Mukherjee:2025ytj, Wang:2025gus, Zhang:2025qbs, Dhawan:2025mer, Kanodia:2025jqh, Li:2025htp, Zheng:2025cgq, Alfano:2025fyq, Avila:2025sjz} for recent differing findings). In contrast, $R_{D_H}(z_{\rm eff})$ exhibits a descending trend that is driven by the highest redshift bin, where the significance of the trend exceeds $2 \sigma$ with Bayesian methods, but reduces to $1.4 \sigma$ with frequentist methods. This confirms that DESI distances and distances reconstructed from DES SNe are consistent at the redshifts probed. Given concerns of systematics, DES SNe have recently been recalibrated through the DES-Dovekie reanalysis \cite{DES:2025sig}. We confirm that replacing DES SNe with DES-Dovekie SNe further improves the consistency. 

We make our analysis more complete by two further studies. It has recently been highlighted \cite{Xu:2026sbw, Pantos:2026rpe} that DESI BAO and transversal BAO disagree on $D_M/r_d$ at $3.7 \sigma$ at $z_{\rm eff} = 0.510$. We find that DES and DES-Dovekie SNe $D_M/r_d$ values are more compatible with DESI BAO than transversal BAO. In the appendix, motivated by differences in the $\Lambda$CDM parameter at effective redshift $z_{\rm eff}$ \cite{Colgain:2025nzf} we compare distances from DESI DR1 BAO and DESI DR1 FS modelling finding good agreement despite small shifts being evident.

\section{Methodology and Data}
Our goal is to crosscheck cosmological distances from DESI DR2 BAO against Type Ia SNe at the same effective redshift. DESI provides direct constraints on the comoving distance $D_M(z)$ and the inverse of the Hubble parameter $D_H(z)$ up to a factor of the radius of the sound horizon $r_d$:
\begin{equation}
\label{eq:def}
    \frac{D_M(z)}{r_d} = \frac{c}{r_d} \int_0^z \frac{1}{H(z^{\prime})} \dd z^{\prime}, \quad \frac{D_H(z)}{r_d} = \frac{c}{r_d H(z)}.  
\end{equation}
In our analysis the most general model we consider is the $w_0 w_a$CDM (CPL) model \cite{Chevallier:2000qy, Linder:2002et}, 
\begin{equation}
\label{eq:Hubble}
    H(z) = H_0 \sqrt{(1-\Omega_m) (1+z)^{3(1+w_0 + w_a)} e^{-\frac{3 w_a z}{1+z}} + \Omega_m (1+z)^3}, 
\end{equation}
with free parameters $H_0, \Omega_m, w_0$ and $w_a$. The model is currently topical as it serves as the basis for dynamical DE claims \cite{DESI:2024mwx, DESI:2024hhd, DESI:2025zgx}. Setting $w_0 = w, w_a=0$ we recover the $w$CDM model, and further restricting $w_0 = -1$, we identify the flat $\Lambda$CDM model. All these models can be viewed as extensions of the minimal $\Lambda$CDM model. 

DESI DR2 BAO constraints \cite{DESI:2025zgx} at effective redshifts $z_{\textrm{eff}}$ are reproduced in Table \ref{tab:BAO}, where $r$ denotes the correlation coefficient between $D_M/r_d$ and $D_H/r_d$ at $z_{\textrm{eff}}$. At lower redshifts, DESI currently only well constrains a combination of $D_M$ and $D_H$ and at $z \gtrsim 1$ SNe samples become sparse, so the crosscheck loses potency. It is worth noting that $D_M$ integrates $D_H$, so BAO and SNe agreeing on $D_M$ does not guarantee agreement on $D_{H}$. Moreover, the integral $D_M$ is effectively a weighted sum, and since $H(z)$ increases ($1/H(z)$ decreases) with $z$, it is less sensitive to high redshift differences in $H(z)$. The distance duality relation \cite{Favale:2024sdq, Teixeira:2025czm, Afroz:2025iwo, Mukherjee:2025ytj, Wang:2025gus, Zhang:2025qbs, Dhawan:2025mer, Kanodia:2025jqh, Li:2025htp, Zheng:2025cgq, Alfano:2025fyq, Avila:2025sjz} reduces to a weaker crosscheck on only $D_M$ distances. 

\begin{table}[H]
\centering
\renewcommand{\arraystretch}{1.3}
\begin{tabular}{|c|c|c|c|}
\hline
$z_{\rm eff}$  & $D_M/r_d$ & $D_H/r_d$ & $r$ \\
\hline\hline
0.510  & $13.588 \pm 0.167$ & $21.863 \pm 0.425$ & $-0.459$ \\
0.706  & $17.351 \pm 0.177$ & $19.455 \pm 0.330$ & $-0.404$\\
0.934  & $21.576 \pm 0.152$ & $17.641 \pm 0.193$ & $-0.416$ \\
\hline
\end{tabular}
\caption{DESI DR2 BAO constraints.}
\label{tab:BAO}
\end{table}

To make comparison, we define the ratios
\begin{equation}
R_{D_M} = \frac{D_M^{\mathrm{SNe}}/r_d}{D_M^{\mathrm{DESI}}/r_d}, 
\qquad
R_{D_H} = \frac{D_H^{\mathrm{SNe}}/r_d}{D_H^{\mathrm{DESI}}/r_d}.
\label{eq:Ratios}
\end{equation}
Although $r_d$ mathematically drops out, we include it to remind the reader that DESI constraints incorporate $r_d$ (Table \ref{tab:BAO}) whereas SNe distances require division by $r_d$.
We note that BAO is calibrated through $r_d$, whereas  SNe are calibrated through the absolute magnitude $M_B$. Therefore, for a given value of $M_B$, there always exists an $r_d$ so that either $R_{D_M} = 1$ or $R_{D_H} =1$. 

To ensure that BAO and SNe are at as close a $z_{\textrm{eff}}$ as possible we consider the standard error $\sigma_k$ weighted redshift for the $k^{\textrm{th}}$ SNe, 
\begin{equation}\label{eq:zeff}
 z_{\textrm{eff}} = \frac{\sum_{k} z_k (\sigma_{k})^{-2}}{\sum_{k} (\sigma_k)^{-2}},  
\end{equation}
where $\sigma_k$ is the distance modulus error. It is worth noting that SNe and galaxies are different observables with different distributions in any common redshift bin. Thus, if one adopted the same redshift bins as DESI, this does not guarantee constraints at the same effective redshift; the only way to do this is to adjust the SNe bins to maintain the same $z_{\rm eff}$ as in Table \ref{tab:BAO}. Conceptually, there is no problem with comparing the constraints on distances from different observables at the same effective redshift. 

One next bins the SNe so that BAO and SNe $z_{\textrm{eff}}$ agree. We employ the following strategy. We construct the highest redshift bin so that it includes the highest redshift SNe and reduce the lower bound on the bin until $z_{\textrm{eff}}$ is close to $z_{\textrm{eff}} = 0.934$ (see Table \ref{tab:BAO}). The lower bound on this bin marks the upper bound on the second bin, thus no SNe are omitted. One proceeds to the lowest redshift bin with $z_{\textrm{eff}} = 0.510$ so that only SNe with redshifts lower than the lowest bin are omitted. The three DES bins are shown in Table \ref{tab:bins}, where there is a negligible difference in $z_{\textrm{eff}}$ in the third bin, confirming that one can construct DES redshift bins with essentially the same $z_{\textrm{eff}}$ as DESI. We stress the uniqueness of the binning strategy. DESI BAO probes higher redshifts than SNe, so it is logical to start from the highest redshifts probed by SNe to construct bins. Moreover, where those bins terminate is dictated by the requirement of a shared effective redshift across the two observables and no SNe are removed between the bins. 

\begin{figure}[H]
\centering
\includegraphics[scale=0.4]{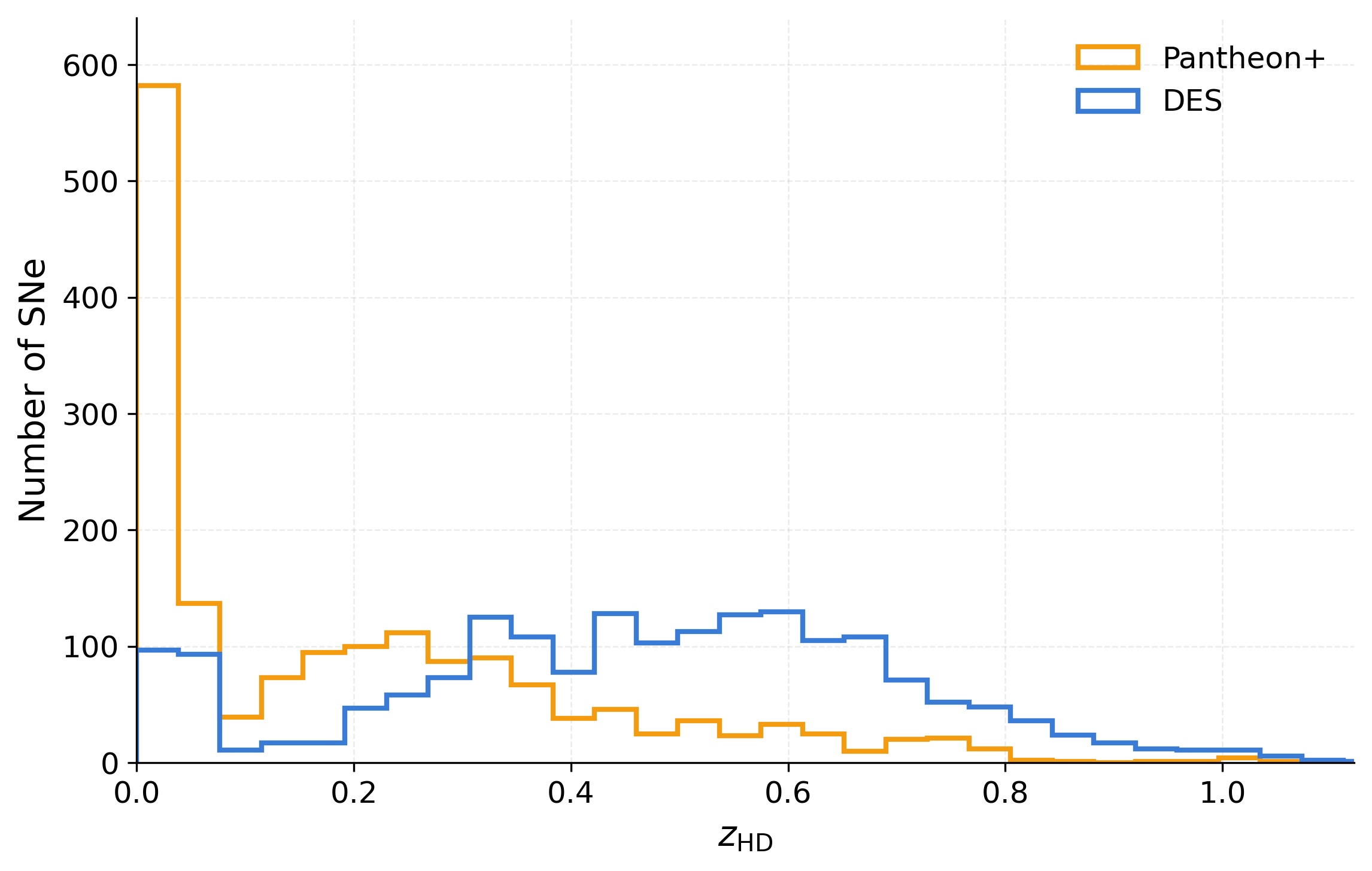}
\caption{Redshift distributions for Pantheon+ (orange) and DES (blue) SNe samples. Pantheon+ contains the majority of its supernovae at low redshift ($z \lesssim 0.2$), while DES provides more uniform coverage up to $z \sim 1$.}
\label{fig:z_distributions}
\end{figure}

This binning strategy singles out the DES sample \cite{DES:2024jxu} over Pantheon+ sample \cite{Scolnic:2021amr, Brout:2022vxf} and Union3 \cite{Rubin:2023ovl} (yet to release full data) as the SNe dataset to give the sternest BAO crosscheck. The advantages of DES are manyfold. First, DES SNe probe higher redshifts, thus ensuring better overlap with DESI BAO. See Fig. \ref{fig:z_distributions} for a visual comparison. Secondly, with DES one works with SNe from a single survey, but systematics may still be present, as evidenced by the recent DES-Dovekie reanalysis \cite{DES:2025sig}. Finally, Pantheon+ has only 30 SNe in the wide range $0.8 < z \leq 2.26$ making it difficult to construct a bin overlapping with the $z_{\textrm{eff}} = 0.934$ DESI constraint.  

\begin{table}[H]
\centering
\renewcommand{\arraystretch}{1.3}
\begin{tabular}{|c| c| c|}
\hline
$z_{\textrm{eff}}$ & \# SNe& $z_{\rm range}$ \\
\hline\hline
0.510 & 748 & $0.3715 < z \leq 0.628$ \\
0.706 & 357 & $0.628 < z \leq 0.826$ \\
0.933 & 101 & $0.826 < z \leq 1.12132$ \\
\hline
\end{tabular}%
\caption{DES redshift bins with number of SNe and effective redshift $z_{\rm eff}$.}
\label{tab:bins}
\end{table}

The task now is to extract a distribution of $D^{\textrm{SNe}}_M/r_d$ and $D^{\textrm{SNe}}_H/r_d$ from DES SNe, so that we can divide it by a normal distribution $\mathcal{N} (Y, \sigma_Y^2)$ where $Y \in \{ D^{\textrm{DESI}}_M/r_d, D^{\textrm{DESI}}_H/r_d\}$. We generate the latter as a bivariate normal distribution with a $2 \times 2$ covariance matrix that incorporates the correlation coefficient $r$. To construct the numerator in the ratios (\ref{eq:Ratios}), we fit the flat $\Lambda$CDM, $w$CDM or $w_0 w_a$CDM model in turn to the binned SNe data with a Gaussian prior on the absolute magnitude $M_B = -19.387 \pm 0.023$ that ensures a $H_0$ consistent with Planck. The $M_B$ prior comes from fitting the $\Lambda$CDM parameters $(H_0, \Omega_m)$ with auxiliary $M_B$ to the full DES SNe sample with a Planck prior $H_0 = 67.36 \pm 0.54$ \cite{Planck:2018vyg}. Note also that in binning the SNe data we truncate the DES covariance matrix.\footnote{This value is $3.8 \sigma$ removed from SH0ES $M_B = -19.253 \pm 0.027$ \cite{Riess:2021jrx}, so the difference in central values encapsulates Hubble tension.} We employ MCMC through \textit{emcee} \cite{Foreman-Mackey:2012any}, and for each entry in the MCMC, we construct $D^{\textrm{SNe}}_M$ and $D^{\textrm{SNe}}_H$ values through  (\ref{eq:def}) and (\ref{eq:Hubble}), leading to a $D^{\textrm{SNe}}_M$ and $D^{\textrm{SNe}}_H$ distribution. The final step is to divide this distribution by a normal distribution $\mathcal{N}(r_d, \sigma_{r_d}^2)$ of equal length to the MCMC chain, before dividing by the $D_M^{\textrm{DESI}}/r_d$ and $D_{H}^{\textrm{DESI}}/r_d$ to get a final distribution for the ratios (\ref{eq:Ratios}). Once we have the final distribution for $R_{D_X}$, $X\in \{M, H \}$ we quote the median ($50^{\textrm{th}}$ percentile) as the central value with $16^{\textrm{th}}$ and $84^{\textrm{th}}$ percentiles marking the extent of the $68 \%$ credible interval.

We also construct frequentist $68 \%$ confidence intervals following the profile likelihood methods of \cite{Gomez-Valent:2022hkb, Colgain:2023bge} (see \cite{Trotta:2017wnx} where the approach is suggested). Concretely, for each entry in our MCMC chain, we extract a $\chi^2$ and reconstruct $D_X(z_{\rm eff})$, $X \in \{M, H\}$ for each $z_{\rm eff}$. We identify the minimum and maximum value of $D_X(z_{\rm eff})$ and divide the interval in between up into 200 odd uniform bins labelled by the value of $\theta = D_{X}(z_{\rm eff})$ at the centre of the bin. In each bin with bin centre $\theta$ we identify the minimum value of $\chi^2$,  $\chi^2_{\rm min}(\theta)$, while we reserve $\chi^2_{\rm min}$ for the global minimum $\chi^2$ across all the bins. By construction, there exists a $\theta$ such that $\chi^2_{\rm min} (\theta) = \chi^2_{\rm min}$. We define the profile likelihood 
\begin{equation}\label{R-theta}
R(\theta) = \exp \left(-\frac{1}{2} (\chi^2_{\textrm{min}}(\theta)-\chi^2_{\textrm{min}}) \right).   
\end{equation}

Since MCMC is not an ideal optimiser, $R(\theta)$ may exhibit some noise, but for converged MCMC chains the agreement between binning the MCMC chain and scanning over parameters while maximising the likelihood shows good agreement as demonstrated in Fig. 4 of \cite{Colgain:2024clf}. The upshot of binning the MCMC chain, as advocated in \cite{Trotta:2017wnx}, is that the construction of Bayesian credible intervals and frequentist confidence intervals start from the same underlying MCMC chain, which sharpens the comparison. Another advantage over scanning and extremising is that one does not need to specify from the beginning the parameter of interest. 

We check that the profile likelihood is Gaussian to a good approximation before applying Wilks' theorem \cite{Wilks:1938dza}, which is valid in the large sample (Gaussian) limit. We symmetrise any difference in the errors by assuming the largest error $\sigma_{D_X}$ before generating a normal distribution $\mathcal{N} (D_{X}, \sigma_{D_X}^2)$, so in our frequentist approach all inputs in (\ref{eq:Ratios}) are normal distributions. Observe that the only difference between Bayesian and frequentist approaches is the construction of the $D_X^{\rm SNe}$ distributions, otherwise the approach is the same. As we shall see in due course, for $R_{D_M}(z_{\rm eff})$ both Bayesian and frequentist methods show excellent agreement with differences arising for $R_{D_H}(z_{\rm eff})$. This essentially confirms that $D_{M}(z_{\rm eff})$ reconstructed from the MCMC chain in a purely Bayesian approach is close to normal.

As a short technical aside, it is standard practice to introduce Gaussian priors on $r_d$. The DESI collaboration adopts this approach in \cite{DESI:2024mwx} to break a degeneracy between $H_0$ and $r_d$. Our setup does not include data that directly constrains the baryonic matter density $\Omega_b$, but if it did, one would find that $r_d$ correlates with $\Omega_m$ through $\Omega_b$. The assumption being made here and in \cite{DESI:2024mwx} is that $r_d$ is uncorrelated from other parameters, which means that more of model parameter space is explored, thereby ensuring that our errors are more conservative. Finally, one expects the error on $r_d$ to increase as one relaxes the $\Lambda$CDM model \cite{Verde:2016wmz} (see also \cite{Vonlanthen:2010cd, Audren:2012wb, Audren:2013nwa, Gomez-Valent:2021hda}). In this sense, the assumption of a Planck-$\Lambda$CDM $r_d$ value is aggressive and relaxing this is expected to lead to greater consistency in a setting where one can already demonstrate consistency within $2 \sigma$. 

In summary, the plan is to reconstruct (\ref{eq:Ratios}) using both Bayesian and frequentist methods under late Universe cosmological model assumptions, which we will now show make little difference to the results. Given that one can always dial $(M_B, r_d)$ to bring $R_{D_M}$ or $R_{D_H}$ to unity, we will primarily be interested in checking that both trace horizontal lines with $z_{\textrm{eff}}$. 

\section{Model Independence}
In this section we expand on our claim that $D_M(z_{\rm eff})$ and $D_H(z_{\rm eff})$ can be reconstructed effectively model independently (agnostically) from SNe in a redshift bin with effective redshift $z_{\rm eff}$. We focus on $z_{\rm eff}$, corresponding to the redshift where the binned data is most constraining. Note, our analysis only covers $z_{\rm eff}$ and we are not claiming that $D_M(z)$ and $D_H(z)$ can be reconstructed model independently across the bin, simply locally at a point. We also remark that our analysis here only covers nested models, more concretely $\Lambda$CDM extendable to the $w_0 w_a$CDM model with two additional parameters, including $w$CDM as a special case, so we do not prove this generically for all models. Note, this setting is sufficient for our purposes of reconstructing SNe distances at the same $z_{\rm eff}$ as BAO. 

Let us begin by reiterating why this should be true. SNe directly constrain luminosity distances proportional to comoving distances, $D_{L}(z) = (1+z) D_{M}(z)$, so when one reconstructs $D_M(z)$ from the model, one should \textit{locally} recover the original constraints. We also note that in a flat universe $D_H(z)$ is simply the derivative of $D_M(z)$,\footnote{With curvature $\Omega_k$, more generally $D_M^{\prime}(z) = D_H(z) \sqrt{1+\Omega_k H_0^2/c^2 D^2_M(z)}$.} so if raw SNe data can directly constrain the functional form of $D_M(z)$, then its first derivative should also be constrained. Throughout, we should bear in mind that the focus in cosmology is usually model parameter space. Although the constraints in model parameter space may look radically different, especially when degeneracies between parameters are at work, our claim here is that these differences are less apparent when $D_M(z)$ and $D_H(z)$ are reconstructed.  

To demonstrate the model independence of distances, we mock up DES SNe data on the $w_0 w_a$CDM cosmology $(H_0, \Omega_m, w_0, w_a) = (63.6, 0.353, -0.42, -1.75)$, corresponding to the DESI+CMB entry in Table V of \cite{DESI:2025zgx}. We do not add any noise, so $\chi^2 = 0$ for $(H_0, \Omega_m, w_0, w_a) = (63.6, 0.353, -0.42, -1.75)$. Our motivation is to pick a relevant $w_0 w_a$CDM cosmology that is distinct from $\Lambda$CDM. We execute the pipeline outlined in the previous section. In Table \ref{tab:expected} we document the expected values to the nearest Mpc based on our injected $w_0 w_a$CDM cosmology. 

\begin{figure}[htb]
\centering
\includegraphics[scale=0.55]{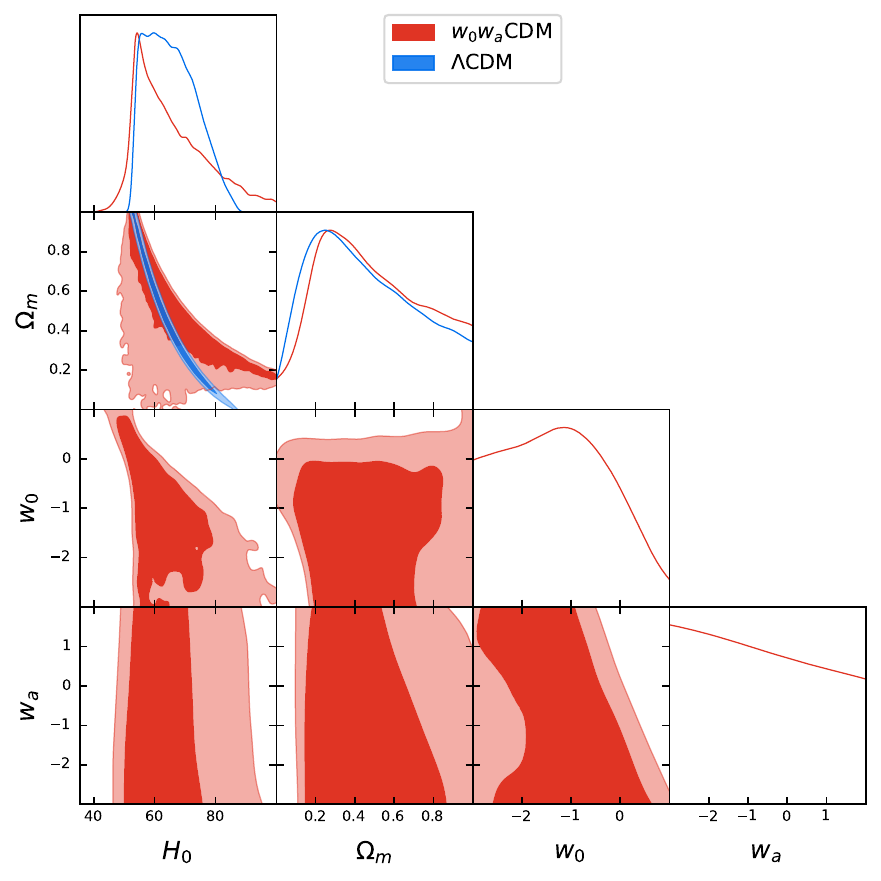}
\caption{Bayesian posteriors for the $\Lambda$CDM and $w_0 w_a$CDM model from highest redshift bin, where degeneracies are present and posteriors are poorly constrained.}
\label{fig:corner}
\end{figure}

In Table \ref{tab:lcdm} and Table \ref{tab:cpl} we present results based purely on Bayesian posteriors for $\Lambda$CDM and $w_0 w_a$CDM, respectively. In Fig. \ref{fig:corner} we present Bayesian posteriors from the highest redshift bin with $z_{\rm eff} = 0.933$ to illustrate poor constraints on the models. In such a setting, there is always a danger of projection effects when one marginalises. In particular, we note the banana-shaped contour in the $(H_0, \Omega_m)$-plane, a hallmark of a poorly broken degeneracy in both models. In our mock analysis in this section, we make use of uniform priors $H_0 \in [0, 100], \Omega_m \in [0, 1], w_0 \in [-3, 1]$ and $w_a \in [-3, 2]$. We remind the reader that each point in model parameter space (each entry in the MCMC chain) is used to reconstruct $D_M(z_{\rm eff})$ and $D_H(z_{\rm eff})$ at each $z_{\rm eff}$ and we quote $16^{\textrm{th}}, 50^{\textrm{th}}$ and $84^{\textrm{th}}$ percentiles for these distances to define our central values and $68 \%$ Bayesian credible intervals.

\begin{table}[H]
\centering
\renewcommand{\arraystretch}{1.3}
\begin{tabular}{|c | c | c|}
\hline
$z_{\textrm{eff}}$ & $D_M$ (Mpc) & $D_H$ (Mpc) \\
\hline\hline
$0.510$ & $1964$ & $3243$ \\
$0.706$ & $2566$ & $2910$ \\
$0.933$ & $3188$ & $2573$ \\
\hline
\end{tabular}%
\caption{Expected constraints for our injected $w_0 w_a$CDM cosmology.}
\label{tab:expected}
\end{table}

\begin{table}[H]
\centering
\renewcommand{\arraystretch}{1.3}
\begin{tabular}{|c |c|c |c | c|}
\hline
$z_{\textrm{eff}}$ & $D_M$ (Mpc) & $\sigma$ & $D_H$ (Mpc) & $\sigma$ \\
\hline\hline
$0.510$ & $1965.1^{+6.5}_{-6.4}$ & $0.17$ & $3226^{+86}_{-87}$ & $0.20$ \\
$0.706$ & $2566\pm 13$ & $0$ &  $2837^{+223}_{-201}$ & $0.33$ \\
$0.933$ & $3187^{+36}_{-37}$ & $0.03$ & $2434^{+342}_{-230}$ & $0.41$ \\
\hline
\end{tabular}%
\caption{Constraints for $D_M$ and $D_H$ reconstructed from the $\Lambda$CDM model with mock $w_0 w_a$CDM data where $\sigma$ denotes the number of standard deviations from the expected values in Table \ref{tab:expected}.}
\label{tab:lcdm}
\end{table}

\begin{table}[H]
\centering
\renewcommand{\arraystretch}{1.3}
\begin{tabular}{|c | c | c| c| c|}
\hline
$z_{\textrm{eff}}$ & $D_M$ (Mpc) & $\sigma$ & $D_H$ (Mpc) & $\sigma$ \\
\hline\hline
$0.510$ & $1965.5^{+6.6}_{-6.5}$ & $0.23$ &  $3223^{+84}_{-87}$ & $0.24$ \\
$0.706$ & $2566 \pm 13$ & $0$ &  $2807^{+230}_{-210}$ & $0.45$ \\
$0.933$ & $3184^{+37}_{-36}$ & $0.11$ &  $2326^{+278}_{-179}$ & $0.89$ \\
\hline
\end{tabular}%
\caption{Same as Table \ref{tab:lcdm} but for $w_0 w_a$CDM.}
\label{tab:cpl}
\end{table}

Referring to Table \ref{tab:lcdm} and Table \ref{tab:cpl}, we see that $D_M(z_{\rm eff})$ agrees well with the expected value across the models. Moreover, the errors do not inflate as we add parameters. Contrast this with Fig. \ref{fig:corner} where it is clear that errors inflate greatly as one extends model parameter space. This reinforces the claimed model independence of $D_M (z_{\rm eff})$. 

However, as explained earlier, while raw SNe data constrain the functional form of $D_L(z) \propto D_M(z)$ well, it may not constrain its derivative $D_H(z)$ as well. That being said, it is clear that the $\Lambda$CDM model tracks the expected $D_H(z_{\rm eff})$ value better than the $w_0 w_a$CDM model and is less biased. It should be stressed here that we have injected a $w_0 w_a$CDM model distinct from $\Lambda$CDM. Moreover, the errors are representative across the models and do not inflate. Across both models we see that the reconstructed $D_H (z_{\rm eff})$ is shifted to lower values than expected across all redshift bins. Since the shift increases with significance at higher redshifts, this risks biasing results based on Bayesian methods. We now explain how this can be corrected. 

\begin{table}[H]
\centering
\renewcommand{\arraystretch}{1.3}
\begin{tabular}{|c |c|c|c|c|}
\hline
& \multicolumn{2}{|c|}{$\Lambda$CDM} & \multicolumn{2}{|c|}{$w_0 w_a$CDM} \\
\hline\hline
$z_{\textrm{eff}}$ & $D_M$ (Mpc) & $D_H$ (Mpc) & $D_M$ (Mpc) & $D_H$ (Mpc)\\
\hline
$0.510$ & $1965.1^{+6.5}_{-6.4}$ & $3243^{+82}_{-87}$ & $1964.0^{+8.6}_{-6.9}$ & $3241^{+86}_{-85}$ \\
$0.706$ & $2567 \pm 13$ & $2912^{+212}_{-214}$ & $2566 \pm 14$ & $2914^{+221}_{-213}$ \\
$0.933$ & $3188^{+38}_{-36}$ & $2572^{+411}_{-410}$ & $3188 \pm 38$ & $2572^{+420}_{-414}$ \\
\hline
\end{tabular}%
\caption{Constraints for $D_M$ and $D_H$ reconstructed from the $\Lambda$CDM and $w_0 w_a$CDM models with mock $w_0 w_a$CDM data using profile likelihoods and Wilks' theorem.}
\label{tab:profile}
\end{table}

\begin{figure}[htb]
\centering
\includegraphics[scale=0.55]{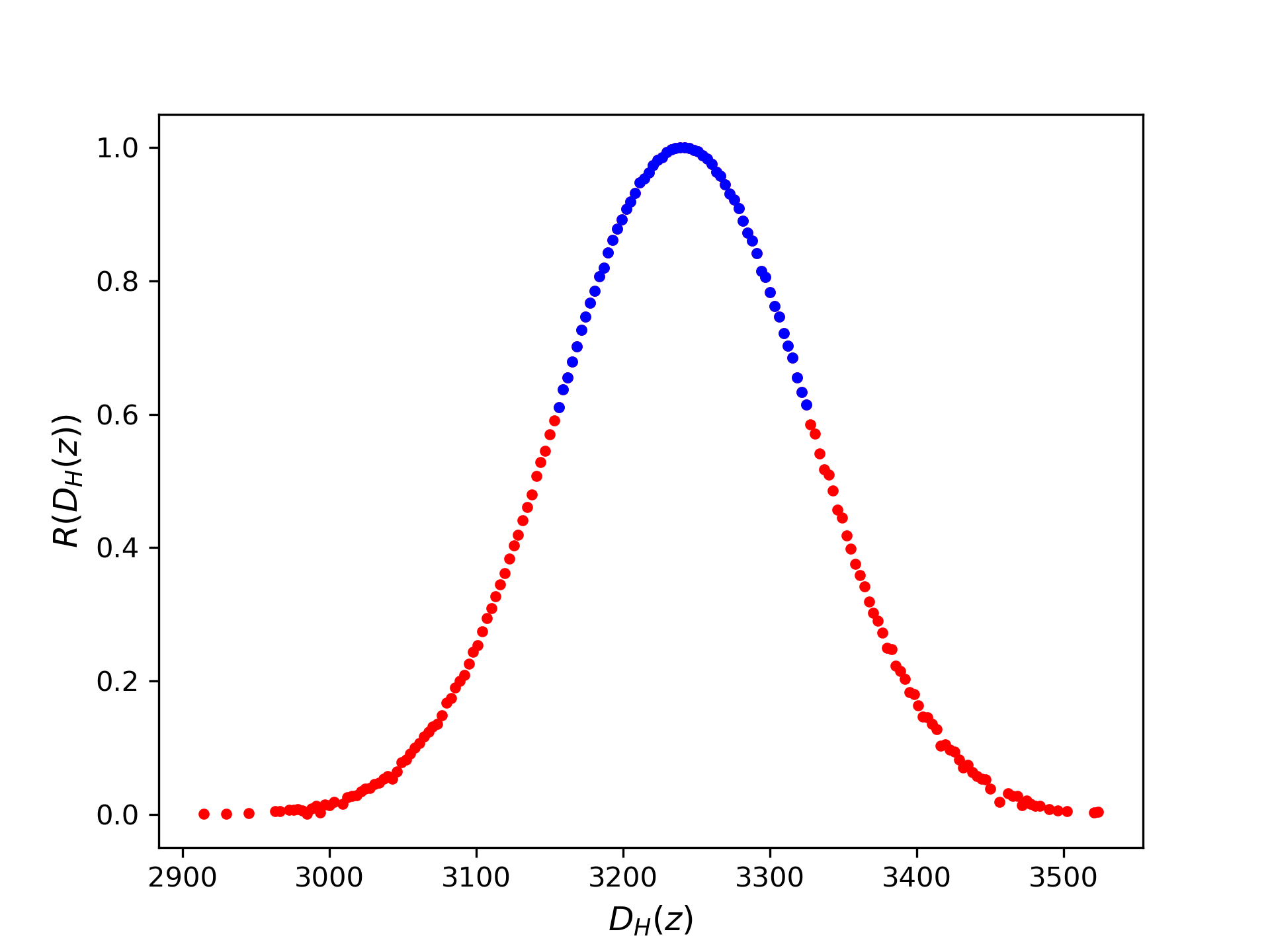}
\caption{The profile likelihood $R(D_H(z))$ reconstructed from the $\Lambda$CDM model at $z_{\rm eff} = 0.510$ with the $68\%$ confidence intervals highlighted in blue. One expects the peak at $D_H(z_{\rm eff}) = 3243$ Mpc and this coincides with the peak of the profile likelihood.
}
\label{fig:profile}
\end{figure}

To check for projection/volume effects in $D_M (z)$ and $D_H (z)$ reconstructed from $\Lambda$CDM and $w_0 w_a$CDM we follow the profile likelihood methods of  \cite{Gomez-Valent:2022hkb, Colgain:2023bge}; see discussion around \eqref{R-theta}. In Table \ref{tab:profile} we use Wilks' theorem \cite{Wilks:1938dza} - the identification of $\Delta \chi^2 \leq 1$ regions in parameter space - to define $68 \%$ frequentist confidence intervals. See Fig. \ref{fig:profile} for an illustrative example of a profile likelihood. What can be seen is that $D_M (z_{\rm eff})$ constraints across both models and both methods show excellent agreement. In contrast, $D_H (z_{\rm eff})$ shifts back to the expected value from the injected mock. A small amount of error inflation is evident across models with profile likelihoods but it is marginal. However, frequentist $68\%$ intervals can vary greatly with Bayesian intervals. What our mock analysis shows is that one can apply either Bayesian or frequentist methods to reconstruct $D_M(z_{\rm eff})$ but $D_H(z_{\rm eff})$ should be reconstructed with purely frequentist methods. Doing so, we can be confident that the reconstructions are effectively model independent for nested models based on $\Lambda$CDM, as we have demonstrated. 

As a further comment, it is evident from Fig. \ref{fig:corner} that in both models data struggles to distinguish $H_0$ and $\Omega_m$ in the highest redshift bin. Such degeneracies impact Bayesian marginalisation and one can expect degeneracies to impact results more as one expands model parameter space from $\Lambda$CDM to $w$CDM to $w_0w_a$CDM. This provides an explanation why the $w_0 w_a$CDM model leads to more biased results than $\Lambda$CDM in our mock analysis.

\section{Results}
Our results with observed DES SNe \cite{DES:2024jxu} can be found in Table \ref{tab:LCDM_Planck} and Table \ref{tab:wCDM_Planck}, where we replace $\Lambda$CDM with $w$CDM to provide a further consistency check on the claimed model independence. We employ uniform priors $H_0 \in [0, 200]$, $\Omega_m \in [0, 4]$ and $w \in [-1.5, -0.5]$ in our MCMC analysis. We relax the traditional $\Omega_m < 1$ prior because DES SNe prefer $\Omega_m > 1$ values at the highest redshifts \cite{Colgain:2024ksa}. Our prior on $w$ may be relatively narrow, but as we have seen earlier in Fig. \ref{fig:corner}, posteriors are poorly constrained in the highest redshift bin even with an injected cosmology with no noise; one expects poorer constraints with observed data. 

\begin{figure}[htb]
\centering
\includegraphics[scale=0.35]{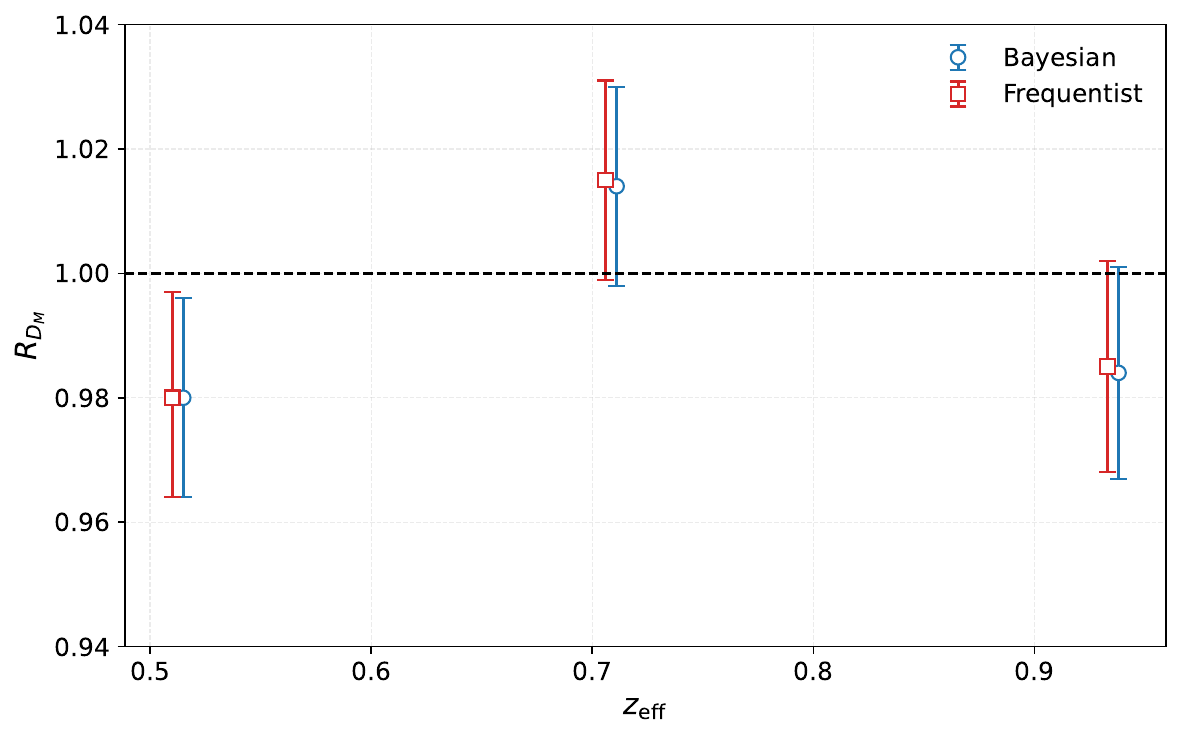}
\caption{$R_{D_M}$ as a function of effective redshift $z_{\textrm{eff}}$ assuming $\Lambda$CDM and $r_d = 147.09 \pm 0.26$ Mpc from Planck. The dashed horizontal line indicates the null hypothesis of consistency, \(R_{D_M} = 1\). Both Bayesian and frequentist methods show excellent agreement.}
\label{fig:RDM_DES}
\end{figure}

With observed data, we confirm that $R_{D_M}(z_{\rm eff})$ leads to the same constraints across Bayesian (MCMC) and frequentist ($\Delta \chi^2 \leq 1$) methods, as expected from our mocks. Fig.~\ref{fig:RDM_DES} confirms that $R_{D_M} (z_{\rm eff})$ scatters around a horizontal. Thus, there exists a choice of $(M_B, r_d)$ that brings one to $R_{D_M} = 1$ and this ensures that the distance duality relation holds (see \cite{Favale:2024sdq, Teixeira:2025czm, Afroz:2025iwo, Mukherjee:2025ytj, Wang:2025gus, Zhang:2025qbs, Dhawan:2025mer, Kanodia:2025jqh, Li:2025htp, Zheng:2025cgq, Alfano:2025fyq, Avila:2025sjz} for discussion). Note, we have not optimised the calibration, simply picked both $M_B$ and $r_d$ consistent with Planck and this led to reasonable results. In contrast, $R_{D_H} (z_{\rm eff})$ exhibits a decreasing trend with increasing $z_{\rm eff}$ that is evident in both Bayesian and frequentist results. This descending trend is primarily driven by the highest redshift bin. 

To unpick this, let us begin with $R_{D_M} (z_{\rm eff})$. As explained earlier, one has the freedom to choose a $(M_B, r_d)$ pair so that $R_{D_M}(z_{\rm eff}) = 1$. We note that irrespective of the methodology, $R_{D_M} (z_{\rm eff})$ is $\sim 1 \sigma $ low at $z_{\rm eff} = 0.510$ and $z_{\rm eff} = 0.934$, and $\sim 1 \sigma$ high at $z_{\rm eff} = 0.706$. In contrast, and restricting our comments to frequentist $\Delta \chi^2 \leq 1$ results, $R_{D_H}(z_{\rm eff})$ is $\sim 1 \sigma$ high at $z_{\rm eff} = 0.510$, consistent with $R_{D_H} (z_{\rm eff}) = 1$ at  $ \sim 0.3 \sigma$ and $\sim 2 \sigma$ low at $z_{\rm eff} = 0.934$. It should be noted that despite the shift in central value between Bayesian and frequentist methods, the errors agree well in the two lowest redshift bins, but in the third bin frequentist errors are larger. A comparison of Table \ref{tab:lcdm}, Table \ref{tab:cpl} and Table \ref{tab:profile} confirms that we observe the same tendency in our mock analysis. Our comments here apply equally to the $w$CDM results in Table \ref{tab:wCDM_Planck}, where the main takeaway message is that allowing an additional model parameter does not change results in a pronounced way, as expected.

\begin{table}[H]
\centering
\renewcommand{\arraystretch}{1.2}
\begin{tabular}{|c | c | c| c |c|}
\hline
& \multicolumn{2}{|c|}{$R_{D_M}$} & \multicolumn{2}{|c|}{$R_{D_H}$} \\
\hline 
$z_{\textrm{eff}}$  & MCMC & $\Delta \chi^2 \leq 1$ & MCMC & $\Delta \chi^2 \leq1 $ \\
\hline \hline
$0.510$ & $0.980 \pm 0.016$ & $0.980^{+0.017}_{-0.016}$ & $1.037^{+0.035}_{-0.034}$ & $1.041^{+0.035}_{-0.034}$ \\
$0.706$ & $1.014 \pm 0.016$ & $1.015 \pm 0.016$ &  $0.994^{+0.080}_{-0.081}$ & $1.025^{+0.080}_{-0.079}$\\
$0.934$ & $0.984 \pm 0.017$ & $0.985 \pm 0.017$ & $0.732^{+0.11}_{-0.056}$ & $0.78 \pm 0.15$ \\
\hline
\end{tabular}%
\caption{$R_{D_M}$ and $R_{D_H}$ at effective redshift $z_{\textrm{eff}}$ assuming $\Lambda$CDM and $r_d = 147.09 \pm 0.26$ Mpc from Planck. We quote both Bayesian (MCMC) and frequentist ($\Delta \chi^2 \leq 1$) results.}
\label{tab:LCDM_Planck}
\end{table}

\begin{table}[H]
\centering
\renewcommand{\arraystretch}{1.2}
\begin{tabular}{|c | c | c| c |c|}
\hline
& \multicolumn{2}{|c|}{$R_{D_M}$} & \multicolumn{2}{|c|}{$R_{D_H}$} \\
\hline 
$z_{\textrm{eff}}$  & MCMC & $\Delta \chi^2 \leq 1$ & MCMC & $\Delta \chi^2 \leq1 $ \\
\hline\hline
$0.510$ & $0.980 \pm 0.016$ & $0.980 \pm 0.016$ & $1.035^{+0.035}_{-0.034}$ & $1.041 \pm 0.036$ \\
$0.706$ & $1.014 \pm 0.016$ & $1.015 \pm 0.016$ & $0.993^{+0.076}_{-0.079}$ & $1.027^{+0.081}_{-0.080}$ \\
$0.934$ & $0.984 \pm 0.017$ & $0.984 \pm 0.017$ &  $0.733^{+0.11}_{-0.055}$ & $0.77 \pm 0.15$ \\
\hline
\end{tabular}%
\caption{Same as Table \ref{tab:LCDM_Planck} but for $w$CDM.}
\label{tab:wCDM_Planck}
\end{table}

\begin{figure}[htb]
\centering
\includegraphics[scale=0.35]{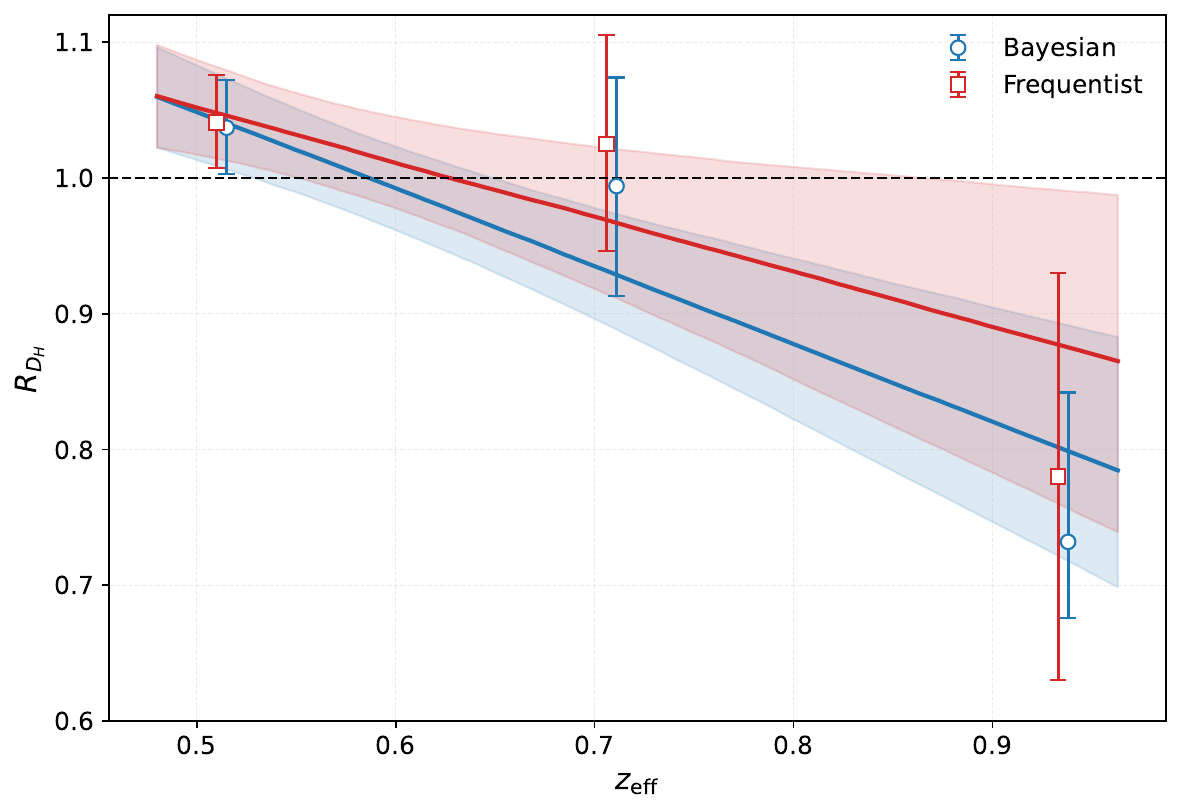}
\caption{$R_{D_H}$ as a function of effective redshift assuming $\Lambda$CDM and $r_d = 147.09 \pm 0.26$ Mpc from Planck. The solid blue and red lines and shaded bands show the best-fit linear trend \(R_{D_H}(z) = m \, z_{\textrm{eff}} + c\) and inferred $68\%$ credible/confidence regions for Bayesian and frequentist methods, while the dashed horizontal line represents the null hypothesis of consistency, \(R_{D_H} = 1\). The lines are $2.5 \sigma$ (blue) and $1.7 \sigma$ (red) removed from horizontal (slope $m=0$).}
\label{fig:RDH_DES}
\end{figure}

Figures \ref{fig:RDM_DES} and \ref{fig:RDH_DES} show that $R_{D_M}(z_{\rm eff})$ and $R_{D_H}(z_{\rm eff})$ trace a horizontal and decreasing trend. Our final exercise is to assess the statistical significance of the descending $R_{D_H} (z_{\rm eff})$ trend. For completeness, we do this for both our Bayesian and frequentist constraints in Table \ref{tab:LCDM_Planck}, where we recall the lessons from the last section that we suspect that a projection/volume effect biases Bayesian results. One could alternatively choose the $w$CDM values in Table \ref{tab:wCDM_Planck}, but the constraints are close so the results cannot change. The approach is to fit a line $y = m x + c$ with slope $m$ and intercept $c$ to the $R_{D_H} (z_{\rm eff})$ constraints on the $y$-axis and $z_{\textrm{eff}}$ on the $x$-axis, while taking into account the difference in errors. We make use of MCMC and define a log-likelihood:   
\begin{equation}
    \ln \mathcal{L}(m, c) = - \frac{1}{2} \sum_{z_{\textrm{eff}}} \frac{[R_{D_H}(z_{\textrm{eff}})-(m \,  z_{\textrm{eff}}+c )]^2}{\sigma^2_{R_{D_H}}}, 
\end{equation}
where if $m \, z_{\textrm{eff}}+c \geq R_{D_H}(z_{\textrm{eff}})$ we select the upper error, and the lower error if not. Referring the reader to Table \ref{tab:LCDM_Planck}, we find $m = -0.58^{+0.23}_{-0.21}$ ($2.5 \sigma$ from the horizontal) with Bayesian constraints and $m = -0.41 \pm 0.29$ ($1.4 \sigma$ from horizontal) with frequentist constraints. Given our preference for frequentist methods, despite the evident decreasing $R_{D_H}(z_{\rm eff})$ trend, we conclude that DES SNe and DESI BAO are consistent within $2 \sigma$. 

We remind the reader again that one is free to employ different assumptions on $r_d$, and by inflating its error in a model agnostic fashion \cite{Verde:2016wmz} one can further decrease the statistical significance of the trend but not in a pronounced way. We omit the details. We also remind the reader again that since $D_M(z)$ integrates $D_H(z)$ and $D_H(z)$ decreases with redshift, $D_M(z)$ is less sensitive to differences in $D_H(z)$, especially as redshift increases. This provides an explanation for why a decreasing trend in $R_{D_H}$ may not be evident in $R_{D_M}$. 

Finally, after the first version of our paper appeared on the arXiv, the DES collaboration conducted a reanalysis of their sample leading to the DES-Dovekie sample \cite{DES:2025sig}. Here we show that this update further marginally reduces the statistical significance of the decreasing $R_{D_H} (z_{\rm eff})$ trend. The biggest difference with our DES analysis of the previous section is that we do not recalibrate the DES-Dovekie sample to make it consistent with the Planck $H_0$ value, which is fixed by the choice of $r_d$. We comment on the difference this makes further below.

From Table \ref{tab:ratios_rd} we note again that the $R_{D_M} (z_{\rm eff})$ constraints show essentially no difference between Bayesian MCMC and frequentist $\Delta \chi^2 \leq 1$ methods. The biggest differences are found in $R_{D_H} (z_{\rm eff})$, where the central values shift upwards for $z_{\rm eff} \in \{0.510, 0.706 \}$ with little difference to the errors, while for $z_{\rm eff} = 0.934$, the central value shifts lower but the errors inflate considerably. Comparing Table \ref{tab:LCDM_Planck} to Table \ref{tab:ratios_rd}, it is clear from both Bayesian and frequentist methods that SNe constraints have improved in the DES-Dovekie reanalysis.   

\begin{table}
\centering
\renewcommand{\arraystretch}{1.2}
\begin{tabular}{|c | c | c| c |c|}
\hline
& \multicolumn{2}{|c|}{$R_{D_M}$} & \multicolumn{2}{|c|}{$R_{D_H}$} \\
\hline 
$z_{\textrm{eff}}$  & MCMC & $\Delta \chi^2 \leq 1$ & MCMC & $\Delta \chi^2 \leq1 $ \\
\hline\hline
0.510 & $0.956 \pm 0.012$ & $0.956 \pm 0.012$ & $1.020^{+0.032}_{-0.031}$ & $1.023^{+0.032}_{-0.031}$ \\
0.706 & $0.981 \pm 0.011$ & $0.981 \pm 0.011$ & $0.986^{+0.076}_{-0.073}$ & $1.013^{+0.074}_{-0.073}$ \\
0.934 & $0.968 \pm 0.011$ & $0.968 \pm 0.012$ & $0.854^{+0.086}_{-0.051}$ & $0.84 \pm 0.12$  \\
\hline
\end{tabular}
\caption{$R_{D_M}$ and $R_{D_H}$ at effective redshift $z_{\textrm{eff}}$ assuming $\Lambda$CDM, DES-Dovekie SNe and $r_d = 147.09 \pm 0.26$ Mpc from Planck. We quote both Bayesian (MCMC) and frequentist ($\Delta \chi^2 \leq 1$) results.}
\label{tab:ratios_rd}
\end{table}

\begin{figure}[htb]
\centering
\includegraphics[scale=0.35]{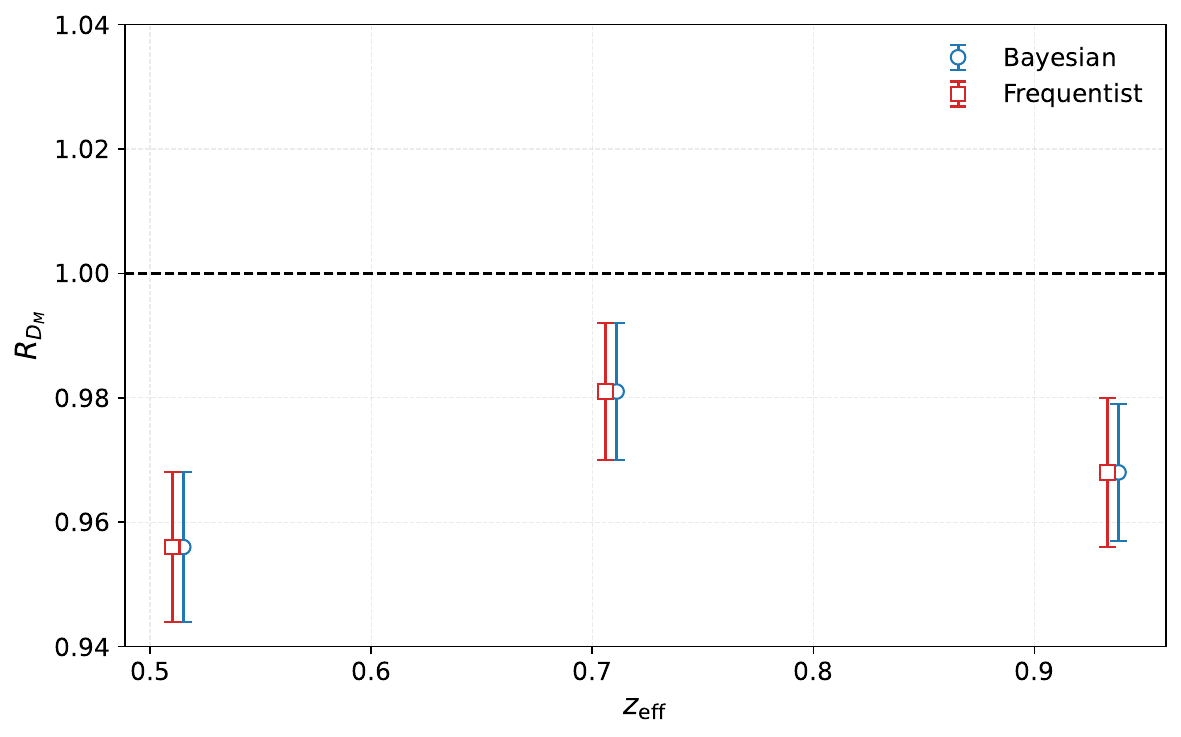}
\caption{Same as Fig. \ref{fig:RDM_DES} but for the DES-Dovekie upgrade. The offset from $R_{D_M} = 1$ is due to DES-Dovekie being calibrated to $H_0 \sim 70$ km/s/Mpc, whereas our $r_d$ calibration is consistent with $H_0 \sim 67.5$ km/s/Mpc.}
\label{fig:RDM_Dovekie}
\end{figure}

\begin{figure}[htb]
\centering
\includegraphics[scale=0.35]{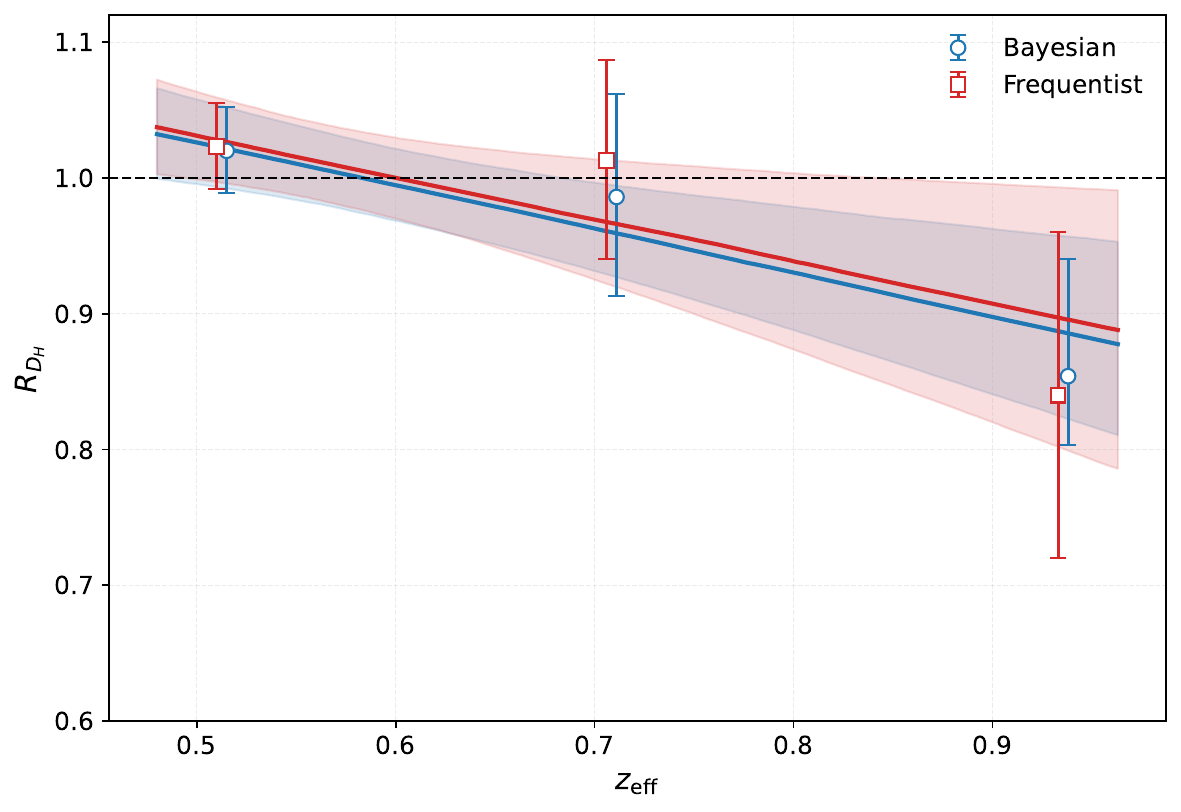}
\caption{Same as Fig. \ref{fig:RDH_DES} but for the DES-Dovekie sample. The blue and red lines are respectively $1.7\sigma$ and $1.2 \sigma$ removed from horizontal (slope $m=0$).}
\label{fig:RDH_Dovekie}
\end{figure}

In Fig. \ref{fig:RDM_Dovekie} it is clear that the $R_{D_M} (z_{\rm eff})$ values in the DES-Dovekie sample continue to trace a horizontal line, but it is noticeably displaced from $R_{D_M}=1$. This is an artifact DES-Dovekie being calibrated to $H_0 \sim 70$ km/s/Mpc and not the Planck value $H_0 \sim 67.5$ km/s/Mpc. In Fig. \ref{fig:RDH_Dovekie} we document the decreasing $R_{D_H}$ trend, where we find a slope of $m = -0.32^{+0.19}_{-0.17}$ with Bayesian methods and $m = -0.31^{+0.25}_{-0.24}$ with frequentist methods. These correspond to a $1.7 \sigma$ and $1.2 \sigma$ effect, respectively. We conclude that DES-Dovekie SNe \cite{DES:2025sig} are more consistent with DESI DR2 BAO than the original DES SNe in the redshifts sampled, where the greatest reduction is seen in Bayesian methods. Given that frequentist methods confirm a significance less than $2 \sigma$ irrespective of the SNe sample, we conclude that SNe distances and DESI BAO are consistent in the redshift ranges studied. Nevertheless, there is a clear descending trend in $R_{D_H}$, which is worth revisiting as both BAO and SNe data improve.

\subsection{Transversal BAO versus DESI BAO}
In \cite{Xu:2026sbw, Pantos:2026rpe} discrepancies between transversal BAO and DESI DR2 BAO constraints are highlighted, where the key point is that the latter are more model ($\Lambda$CDM) dependent. The observation \cite{Pantos:2026rpe} is that transversal BAO prefer lower $D_M/r_d$ values. The discrepancy is most pronounced $z_{\rm eff} = 0.510$ where DESI reports $D_M/r_d = 13.588 \pm 0.167$ but transversal BAO leads to $D_M/r_d=11.912\pm 0.421$ \cite{Carvalho:2015ica}, corresponding to a $3.7 \sigma $ discrepancy at the same effective redshift. This begs the question, which probe is more consistent with SNe?

Since we have calculated $R_{D_M} (z_{\rm eff})$ it is easy to quote intermediate results that get suppressed in the final result. For DES and DES-Dovekie we find $D_M/r_d = 13.32 \pm 0.15$ and $D_M/r_d = 12.983 \pm 0.048$ respectively, where it should be borne in mind that we use the Planck $r_d$ value. As highlighted earlier, the lower DES-Dovekie result follows from a $H_0 \sim 70$ km/s/Mpc calibration versus $H_0 \sim 67.5$ km/s/Mpc, which explains the offset from $R_{D_M} = 1$ in Fig. \ref{fig:RDM_Dovekie}. Since Fig. \ref{fig:RDM_DES} shows good agreement with $R_{D_M} = 1$, we can estimate the $r_d$ required to restore consistency by assuming a constant combination $H_0 r_d$, which means one should rescale the Planck $r_d$ value to $r_d = 142$ Mpc where we assume that the $r_d$ error is unaffected. This shifts the DES-Dovekie constraint upwards to $13.468 \pm 0.051$, which is consistent with DES at $0.9 \sigma$. With this correction, the DES and DES-Dovekie constraints are consistent with DESI BAO at $1.2 \sigma$ and $0.7 \sigma$, while being inconsistent with transversal BAO at $3.2 \sigma$ and $3.7 \sigma$, respectively. 

The main takeaway is that both DES and DES-Dovekie SNe prefer DESI BAO values. Although it is plausible that this arises from assumptions on a fiducial model in both observables, it is more likely that this arises from systematics in transversal BAO. We also emphasise that $H_0$ tension cannot change our results since $D_M/r_d \propto 1/(H_0 r_d)$ and proposals to reduce the radius of the sound horizon $r_d$ in order to increase $H_0$, while retaining constant $H_0 r_d$, do not affect $D_M/r_d$. If systematics are not at play, assuming $E(z) \equiv H(z)/H_0$ does not change greatly, there appears to be a disagreement in $H_0 r_d$.

\section{Discussion}
BAO gives us direct constraints on $D_M/r_d$ and $D_H/r_d$ at set effective redshifts. In this work, we binned DES and DES-Dovekie SNe samples \cite{DES:2024jxu, DES:2025sig} so that the bins possess as close as possible to the same effective redshift, and inferred the analogous constraints. The scaffolding to do this is provided by the $\Lambda$CDM model, but we have argued and demonstrated through mock analysis that $\Lambda$CDM can be replaced with an extended model to arrive at similar results. We constructed ratios $R_{D_M}$ and $R_{D_H}$ noting that $R_{D_M}$ is consistent with a horizontal, implying that there is a choice of calibrating pair $(M_B, r_d)$ so that $R_{D_M} =1$. This implies that any problem with the distance duality relation stems from calibration. 

On the contrary, $R_{D_H}$ exhibits a decreasing trend with effective redshift, the statistical significance of which depends on whether one employs Bayesian or frequentist methods. Our mock analysis confirms a bias in Bayesian methods, which we trace to projection or volume effects, providing a reason to place less emphasis on Bayesian results. Our frequentist results show that both DES and DES-Dovekie lead to a decreasing $R_{D_H}$ trend that is consistent with a horizontal within $2 \sigma$. Thus, we conclude that DES/DES-Dovekie and DESI distances at $z_{\rm} \in \{0.510, 0.706, 0.934\}$ are consistent, with DES-Dovekie distances being more consistent. Nevertheless, both Bayesian and frequentist methods agree that $R_{D_H}$ decreases, making it imperative to revisit these tests as data improve.  

Distinguishing the dataset that gives rise to the decreasing $R_{D_H}$ trend is tricky. On the contrary, it is much easier to single out the $z_{\rm eff} = 0.934$ bin that drives the trend. The fact that $R_{D_M}$ scatters about $R_{D_M}=1$ in Fig. \ref{fig:RDM_DES}, implying a consistent $(M_B, r_d)$ calibrating pair, while the third bin is most displaced from $R_{D_H} = 1$ in Fig. \ref{fig:RDH_DES}, suggests that the $z_{\rm eff} = 0.934$ bin is the outlier. We remind the reader that Pantheon+ \cite{Brout:2022vxf} becomes sparse at higher redshifts and Union3 \cite{Rubin:2023ovl} has yet to officially release data, so DES and DES-Dovekie are the optimal SNe datasets. 

Our Bayesian findings with the DES sample agree with \cite{Dinda:2025hiu} that there may be a high redshift anomaly (claimed to be at $\gtrsim 3 \sigma$ significance in \cite{Dinda:2025hiu}), but since our frequentist results do not back this up, it suggests that any $> 2 \sigma$ discrepancy is driven by methodology. We note that \cite{Dinda:2025hiu} employs Gaussian Procees regression \cite{Holsclaw:2010nb, Holsclaw:2010sk, Shafieloo:2012ht, Seikel:2012uu}, where there is invariably hidden modelling assumptions \cite{OColgain:2021pyh}, whereas here our analysis is based on familiar and trusted parametric models and we have explicitly demonstrated that extending the model makes little difference.  Nevertheless, we agree with \cite{Dinda:2026uff} that DES-Dovekie SNe are more consistent with DESI BAO than DES SNe. Interestingly, there has been considerable discussion on a low redshift offset in SNe explaining DESI's dynamical DE signal \cite{Efstathiou:2024xcq, Notari:2024zmi, Huang:2025som} (see also earlier \cite{Huang:2024qno, Gialamas:2024lyw}), but there are hints of disagreements between DESI BAO and DES SNe at higher redshifts, admittedly at low statistical significance. It cannot be stressed enough that it is prudent science to check that different observables agree on distances either to the same galaxy, e. g. \cite{Li:2024pjo}, or at the same effective redshift, before attempting to combine datasets. These crosschecks should go hand in hand with big physical claims.

Finally, as a byproduct of our analysis, we are able to confirm that DES and DES-Dovekie SNe imply that $D_M/r_d$ values at $z_{\rm eff} = 0.510$ are more consistent with DESI DR2 BAO \cite{DESI:2025zgx} than transversal BAO \cite{Carvalho:2015ica}. This cannot be naively traced to $H_0$ tension, and neglecting a discrepancy in $H_0 r_d$, the most likely explanation is systematics in transversal BAO.

\section*{Acknowledgements} 
We thank Ioannis Pantos and Leandros Perivolaropoulos for correspondence. This article/publication is based upon work from COST Action CA21136 – “Addressing observational tensions in cosmology with systematics and fundamental physics (CosmoVerse)”, supported by COST (European Cooperation in Science and Technology). MLH is supported by Secihti grant 806098. The work of MMShJ is in part supported by the INSF research chair grant No.40405163.

\appendix 

\section{A check on BAO and FS modeling}
Here we check whether constraints on $D_M/r_d, D_H/r_d$ and $D_V/r_d := (z D_M^2 D_H)^{\frac{1}{3}}/r_d$ from DESI DR1 BAO \cite{DESI:2024mwx} and DESI DR1 FS modelling \cite{DESI:2024hhd, DESI:2024jxi} are consistent. Our interest is driven by mild differences in the inferred $\Lambda$CDM parameter $\Omega_m$ from these methods for luminous red galaxy (LRG) data at $z_{\textrm{eff}} = 0.51$ and $z_{\textrm{eff}} = 0.706$. The differences can be seen from Fig. 5 of \cite{Colgain:2025nzf}, where the $\Omega_m$ difference for $z_{\textrm{eff}} = 0.51$ LRG is most pronounced. The likely origin of these discrepancies is statistical fluctuations in BAO data \cite{DESI:2024mwx, Colgain:2024xqj}. The main lesson of this section is that shifts in the cosmological parameter $\Omega_m$ between BAO and FS modelling highlighted in \cite{Colgain:2024mtg, Colgain:2025nzf} can be seen directly in the distances. 

The results for BAO can be found in Table 1 of \cite{DESI:2024mwx}, which we reproduce in Table \ref{tab:BAO_constraints}. For FS modelling, we can reconstruct the quantities from the $(H_0, \Omega_m)$ values for FS modelling alone (no BAO) in Table 10 of \cite{DESI:2024jxi}. The analogous Lyman-$\alpha$ results, which are omitted in \cite{DESI:2024jxi}, can be found in \cite{Cuceu:2025nvl}. It should be stress that $(H_0, \Omega_m)$ are correlated, but here we treat them as uncorrelated parameters. This leads to overestimated errors making our analysis more conservative. As a result, the statistical significance of the shifts are bounded below by our analysis, thus making them more serious. The existence of the direct $(D_M/r_d, D_H/r_d)$ constraints corresponding to Table 10 of \cite{DESI:2024jxi} in the literature would allow a better comparison.     

\begin{table}[htb]
\centering
\renewcommand{\arraystretch}{1.3}
\begin{tabular}{|c | c | c|c|}
\hline
$z_{\textrm{eff}}$ & $D_V/r_d$  & $D_M/r_d$ & $D_H/r_d$  \\
\hline\hline
$0.295$ & $7.93 \pm 0.15$ & $-$ & $-$ \\
$0.510$ & $-$ & $13.62\pm 0.25$ & $20.98 \pm 0.61$ \\
$0.706$ & $-$ & $16.85 \pm 0.32$ & $20.08 \pm 0.60$ \\
$0.930$ & $-$ & $21.71 \pm 0.28$ & $17.88 \pm 0.35$ \\
$1.317$ & $-$ & $27.79 \pm 0.69$ & $13.82 \pm 0.42$ \\
$1.491$ & $26.07 \pm 0.67$ & $-$ & $-$ \\
$2.33$ & $-$ & $39.71 \pm 0.94$ & $8.52 \pm 0.17$ \\
\hline
\end{tabular}%
\caption{DESI DR1 BAO constraints on $D_V/r_d$, $D_M/r_d$ and $D_H/r_d$ at effective redshift $z_{\textrm{eff}}$.}
\label{tab:BAO_constraints}
\end{table}

A further subtlety we handle is the asymmetric errors on $(H_0, \Omega_m)$ in Table 10 \cite{DESI:2024jxi}. We model this by gluing together normal distributions with different standard deviations. This guarantees that the $16^{\textrm{th}}$, $50^{\textrm{th}}$ (median) and $84^{\textrm{th}}$ percentiles recover the central values and the asymmetric errors, $X = A^{+B}_{-C}$. To get an appropriate distribution of $X$ values, we generate two arrays of $4 \times 10^5$ values in normal distributions $\mathcal{N}(A, B^2)$ and $\mathcal{N}(A, C^2)$ and throw away all values less and greater than $A$, respectively. We then trim the remaining values to $1.5 \times 10^5$ and concatenate the two arrays to form a distribution of $3 \times 10^5$ $X \in \{H_0, \Omega_m \}$ values with the appropriate percentiles by construction. Once we have an array of of $3 \times 10^5$ $(H_0, \Omega_m)$ pairs, we construct $D_V, D_M$ and/or $D_H$ at the relevant effective redshift $z_{\textrm{eff}}$ for each pair. Finally, we generate $3 \times 10^5$ Planck $r_d$ values in a normal distribution $\mathcal{N}(r_d, \sigma_{r_d}^2) $, where $r_d = 147.09 \pm 0.26$ \cite{Planck:2018vyg}, and divide the distance arrays $D_V, D_M$ and $D_H$ through by the $r_d$ array, before extracting $16^{\textrm{th}}, 50^{\textrm{th}}$ and $84^{\textrm{th}}$ percentiles. The result of this exercise are shown in Table \ref{tab:FS_constraints}. 

\begin{table}[H]
\centering
\renewcommand{\arraystretch}{1.3}
\begin{tabular}{|c | c | c| c|}
\hline
$z_{\textrm{eff}}$ & $D_V/r_d$  & $D_M/r_d$ & $D_H/r_d$  \\
\hline\hline
$0.295$ & $7.83^{+0.44}_{-0.37}$ & $-$ & $-$ \\
$0.510$ & $-$ & $13.13^{+0.51}_{-0.43}$ & $22.26^{+1.0}_{-0.87}$ \\
$0.706$ & $-$ & $16.58^{+0.65}_{-0.61}$ & $19.19^{+0.94}_{-0.86}$ \\
$0.919$ & $-$ & $21.78^{+0.88}_{-0.82}$ & $17.97^{+0.97}_{-0.88}$ \\
$1.317$ & $-$ & $27.2^{+1.9}_{-1.5}$ & $13.85^{+1.3}_{-0.97}$ \\
$1.491$ & $25.2^{+2.3}_{-1.7}$ & $-$ & $-$ \\
$2.33$ & $-$ & $39.21^{+2.2}_{-1.9}$ & $8.65^{+0.65}_{-0.52}$ \\
\hline
\end{tabular}
\caption{Constraints on $D_V/r_d$, $D_M/r_d$ and $D_H/r_d$ inferred from DESI assuming the $\Lambda$CDM model and the Planck value $r_d = 147.09 \pm 0.26$ Mpc.}
\label{tab:FS_constraints}
\end{table}

We can now compare Table \ref{tab:BAO_constraints} and Table \ref{tab:FS_constraints}. The biggest differences occur for LRG data at $z_{\textrm{eff}} = 0.51$ and $z_{\textrm{eff}} = 0.706$. Note also that the effective redshifts $z_{\textrm{eff}} = 0.919$ and $z_{\textrm{eff}} = 0.930$ disagree, but this is explained by a difference in the tracers. At $z_{\textrm{eff}} = 0.51$, the $D_M/r_d$ and $D_H/r_d$ values are shifted by $0.9 \sigma$ and $1.2 \sigma$ between BAO and FS modelling. At $z_{\textrm{eff}}=0.706$, $D_M/r_d$ and $D_H/r_d$ are shifted by $0.4 \sigma$ and $0.8 \sigma$, respectively. These are the most pronounced shifts we observe and all other tracers show better agreement.\footnote{Note that errors are overestimated by our method. One can compare our results for Lyman-$\alpha$ at $z_{\textrm{eff}}$ to the constraints, $D_M/r_d = 39.05 \pm 0.52$ and $D_H/r_d = 8.63 \pm 0.11$ \cite{Cuceu:2025nvl}.} On the whole this is the expected outcome. There is a more dramatic shift in $(D_M/r_d, D_H/r_d)$ between BAO and FS modelling for LRG and this correlates with differences in $\Omega_m$ seen in Fig. 5 of \cite{Colgain:2025nzf}. We expect these differences to decrease as more data is collected.

\bibliography{refs}

@article{Colgain:2024clf,
    author = "\'O Colg{\'a}in, Eoin and Sheikh-Jabbari, M. M. and Yin, Lu",
    title = "{Do high redshift QSOs and GRBs corroborate JWST?}",
    eprint = "2405.19953",
    archivePrefix = "arXiv",
    primaryClass = "astro-ph.CO",
    doi = "10.1016/j.dark.2025.101975",
    journal = "Phys. Dark Univ.",
    volume = "49",
    pages = "101975",
    year = "2025"
}

@article{Carvalho:2015ica,
    author = "Carvalho, G. C. and Bernui, A. and Benetti, M. and Carvalho, J. C. and Alcaniz, J. S.",
    title = "{Baryon Acoustic Oscillations from the SDSS DR10 galaxies angular correlation function}",
    eprint = "1507.08972",
    archivePrefix = "arXiv",
    primaryClass = "astro-ph.CO",
    doi = "10.1103/PhysRevD.93.023530",
    journal = "Phys. Rev. D",
    volume = "93",
    number = "2",
    pages = "023530",
    year = "2016"
}

@article{DES:2024jxu,
    author = "Abbott, T. M. C. and others",
    collaboration = "DES",
    title = "{The Dark Energy Survey: Cosmology Results with \ensuremath{\sim}1500 New High-redshift Type Ia Supernovae Using the Full 5 yr Data Set}",
    eprint = "2401.02929",
    archivePrefix = "arXiv",
    primaryClass = "astro-ph.CO",
    reportNumber = "FERMILAB-PUB-23-0821-PPD, DES-2023-805",
    doi = "10.3847/2041-8213/ad6f9f",
    journal = "Astrophys. J. Lett.",
    volume = "973",
    number = "1",
    pages = "L14",
    year = "2024"
}

@article{Pantos:2026rpe,
    author = "Pantos, Ioannis and Perivolaropoulos, Leandros",
    title = "{On the origin of the BAOtr-DESI tension}",
    eprint = "2604.11106",
    archivePrefix = "arXiv",
    primaryClass = "astro-ph.CO",
    month = "4",
    year = "2026"
}

@article{Xu:2026sbw,
    author = {Xu, Tengpeng and Kumar, Suresh and Chen, Yun and Capistrano, Abra{\~a}o J. S. and Akarsu, {\"O}zg{\"u}r},
    title = "{Probing Dynamical Dark Energy with Late-Time Data: Evidence, Tensions, and the Limits of the $w_0w_a$CDM Framework}",
    eprint = "2602.11936",
    archivePrefix = "arXiv",
    primaryClass = "astro-ph.CO",
    month = "2",
    year = "2026"
}

@article{RoyChoudhury:2024wri,
    author = "Roy Choudhury, Shouvik and Okumura, Teppei",
    title = "{Updated Cosmological Constraints in Extended Parameter Space with Planck PR4, DESI Baryon Acoustic Oscillations, and Supernovae: Dynamical Dark Energy, Neutrino Masses, Lensing Anomaly, and the Hubble Tension}",
    eprint = "2409.13022",
    archivePrefix = "arXiv",
    primaryClass = "astro-ph.CO",
    doi = "10.3847/2041-8213/ad8c26",
    journal = "Astrophys. J. Lett.",
    volume = "976",
    number = "1",
    pages = "L11",
    year = "2024"
}

@article{Foreman-Mackey:2012any,
    author = "Foreman-Mackey, Daniel and Hogg, David W. and Lang, Dustin and Goodman, Jonathan",
    title = "{emcee: The MCMC Hammer}",
    eprint = "1202.3665",
    archivePrefix = "arXiv",
    primaryClass = "astro-ph.IM",
    doi = "10.1086/670067",
    journal = "Publ. Astron. Soc. Pac.",
    volume = "125",
    pages = "306--312",
    year = "2013"
}

@article{Dinda:2026uff,
    author = "Dinda, Bikash R. and Maartens, Roy and Clarkson, Chris",
    title = "{Calibration-independent consistency test of BAO and SNIa data: update}",
    eprint = "2601.16229",
    archivePrefix = "arXiv",
    primaryClass = "astro-ph.CO",
    month = "1",
    year = "2026"
}

@article{DES:2025sig,
    author = "Popovic, B. and others",
    collaboration = "DES",
    title = "{The Dark Energy Survey Supernova Program: A Reanalysis Of Cosmology Results And Evidence For Evolving Dark Energy With An Updated Type Ia Supernova Calibration}",
    eprint = "2511.07517",
    archivePrefix = "arXiv",
    primaryClass = "astro-ph.CO",
    reportNumber = "FERMILAB-PUB-25-0842-CSAID-PPD",
    month = "11",
    year = "2025"
}

@article{RoyChoudhury:2025dhe,
    author = "Roy Choudhury, Shouvik",
    title = "{Cosmology in Extended Parameter Space with DESI Data Release 2 Baryon Acoustic Oscillations: A 2{\ensuremath{\sigma}}+ Detection of Nonzero Neutrino Masses with an Update on Dynamical Dark Energy and Lensing Anomaly}",
    eprint = "2504.15340",
    archivePrefix = "arXiv",
    primaryClass = "astro-ph.CO",
    doi = "10.3847/2041-8213/ade1cc",
    journal = "Astrophys. J. Lett.",
    volume = "986",
    number = "2",
    pages = "L31",
    year = "2025"
}

@article{Dinda:2025hiu,
    author = "Dinda, Bikash R. and Maartens, Roy and Clarkson, Chris",
    title = "{Calibration-independent consistency test of DESI DR2 BAO and SNIa}",
    eprint = "2509.19899",
    archivePrefix = "arXiv",
    primaryClass = "astro-ph.CO",
    month = "9",
    year = "2025"
}

@article{CosmoVerseNetwork:2025alb,
    author = "Di Valentino, Eleonora and others",
    collaboration = "CosmoVerse Network",
    title = "{The CosmoVerse White Paper: Addressing observational tensions in cosmology with systematics and fundamental physics}",
    eprint = "2504.01669",
    archivePrefix = "arXiv",
    primaryClass = "astro-ph.CO",
    doi = "10.1016/j.dark.2025.101965",
    journal = "Phys. Dark Univ.",
    volume = "49",
    pages = "101965",
    year = "2025"
}

@article{Colgain:2025fct,
    author = "\'O Colg{\'a}in, Eoin and Pourojaghi, Saeed and Sheikh-Jabbari, M. M.",
    title = "{On the Analysis Dependence of DESI Dynamical Dark Energy}",
    eprint = "2505.19029",
    archivePrefix = "arXiv",
    primaryClass = "astro-ph.CO",
    doi = "10.3390/galaxies13060133",
    journal = "Galaxies",
    volume = "13",
    number = "6",
    pages = "133",
    year = "2025"
}

@article{Li:2024pjo,
    author = "Li, Siyang and Anand, Gagandeep S. and Riess, Adam G. and Casertano, Stefano and Yuan, Wenlong and Breuval, Louise and Macri, Lucas M. and Scolnic, Daniel and Beaton, Rachael and Anderson, Richard I.",
    title = "{Tip of the Red Giant Branch Distances with JWST. II. I-band Measurements in a Sample of Hosts of 10 Type Ia Supernova Match HST Cepheids}",
    eprint = "2408.00065",
    archivePrefix = "arXiv",
    primaryClass = "astro-ph.CO",
    doi = "10.3847/1538-4357/ad84f3",
    journal = "Astrophys. J.",
    volume = "976",
    number = "2",
    pages = "177",
    year = "2024"
}

@article{Cuceu:2025nvl,
    author = "Cuceu, Andrei and others",
    title = "{DESI DR1 Ly$α$ forest: 3D full-shape analysis and cosmological constraints}",
    eprint = "2509.15308",
    archivePrefix = "arXiv",
    primaryClass = "astro-ph.CO",
    reportNumber = "FERMILAB-PUB-25-0693-PPD",
    month = "9",
    year = "2025"
}

@article{DESI:2024jxi,
    author = "Adame, A. G. and others",
    collaboration = "DESI",
    title = "{DESI 2024 V: Full-Shape galaxy clustering from galaxies and quasars}",
    eprint = "2411.12021",
    archivePrefix = "arXiv",
    primaryClass = "astro-ph.CO",
    reportNumber = "FERMILAB-PUB-24-0847-PPD",
    doi = "10.1088/1475-7516/2025/09/008",
    journal = "JCAP",
    volume = "09",
    pages = "008",
    year = "2025"
}

@article{Vonlanthen:2010cd,
    author = {Vonlanthen, Marc and R{\"a}s{\"a}nen, Syksy and Durrer, Ruth},
    title = "{Model-independent cosmological constraints from the CMB}",
    eprint = "1003.0810",
    archivePrefix = "arXiv",
    primaryClass = "astro-ph.CO",
    reportNumber = "CERN-PH-TH-2010-053",
    doi = "10.1088/1475-7516/2010/08/023",
    journal = "JCAP",
    volume = "08",
    pages = "023",
    year = "2010"
}

@article{Audren:2013nwa,
    author = "Audren, Benjamin",
    title = "{Separate Constraints on Early and Late Cosmology}",
    eprint = "1312.5696",
    archivePrefix = "arXiv",
    primaryClass = "astro-ph.CO",
    doi = "10.1093/mnras/stu1457",
    journal = "Mon. Not. Roy. Astron. Soc.",
    volume = "444",
    number = "1",
    pages = "827--832",
    year = "2014"
}

@article{Audren:2012wb,
    author = "Audren, Benjamin and Lesgourgues, Julien and Benabed, Karim and Prunet, Simon",
    title = "{Conservative Constraints on Early Cosmology: an illustration of the Monte Python cosmological parameter inference code}",
    eprint = "1210.7183",
    archivePrefix = "arXiv",
    primaryClass = "astro-ph.CO",
    reportNumber = "CERN-PH-TH-2012-290, LAPTH-048-12",
    doi = "10.1088/1475-7516/2013/02/001",
    journal = "JCAP",
    volume = "02",
    pages = "001",
    year = "2013"
}

@article{Verde:2016wmz,
    author = "Verde, Licia and Bellini, Emilio and Pigozzo, Cassio and Heavens, Alan F. and Jimenez, Raul",
    title = "{Early Cosmology Constrained}",
    eprint = "1611.00376",
    archivePrefix = "arXiv",
    primaryClass = "astro-ph.CO",
    doi = "10.1088/1475-7516/2017/04/023",
    journal = "JCAP",
    volume = "04",
    pages = "023",
    year = "2017"
}

@article{Gomez-Valent:2021hda,
    author = "G{\'o}mez-Valent, Adri{\`a}",
    title = "{Measuring the sound horizon and absolute magnitude of SNIa by maximizing the consistency between low-redshift data sets}",
    eprint = "2111.15450",
    archivePrefix = "arXiv",
    primaryClass = "astro-ph.CO",
    doi = "10.1103/PhysRevD.105.043528",
    journal = "Phys. Rev. D",
    volume = "105",
    number = "4",
    pages = "043528",
    year = "2022"
}

@article{DESI:2025zgx,
    author = "Abdul Karim, M. and others",
    collaboration = "DESI",
    title = "{DESI DR2 Results II: Measurements of Baryon Acoustic Oscillations and Cosmological Constraints}",
    eprint = "2503.14738",
    archivePrefix = "arXiv",
    primaryClass = "astro-ph.CO",
    reportNumber = "FERMILAB-PUB-25-0169-PPD",
    month = "3",
    year = "2025"
}

@article{Colgain:2025nzf,
    author = "\'O Colg\'ain, Eoin and Pourojaghi, Saeed and Sheikh-Jabbari, M. M. and Yin, Lu",
    title = "{How much has DESI dark energy evolved since DR1?}",
    eprint = "2504.04417",
    archivePrefix = "arXiv",
    primaryClass = "astro-ph.CO",
    month = "4",
    year = "2025"
}

@article{Colgain:2024mtg,
    author = {\'O Colg\'ain, Eoin and Sheikh-Jabbari, M. M.},
    title = {DESI and SNe: dynamical dark energy, Ωm tension or systematics?},
    journal = {Monthly Notices of the Royal Astronomical Society: Letters},
    volume = {542},
    number = {1},
    pages = {L24-L30},
    year = {2025},
    month = {06},
    issn = {1745-3925},
    doi = {10.1093/mnrasl/slaf042},
    url = {https://doi.org/10.1093/mnrasl/slaf042},
    eprint = "2412.12905",
    archivePrefix = "arXiv",
    primaryClass = "astro-ph.CO",
}

@article{Brout:2022vxf,
    author = "Brout, Dillon and others",
    title = "{The Pantheon+ Analysis: Cosmological Constraints}",
    eprint = "2202.04077",
    archivePrefix = "arXiv",
    primaryClass = "astro-ph.CO",
    doi = "10.3847/1538-4357/ac8e04",
    journal = "Astrophys. J.",
    volume = "938",
    number = "2",
    pages = "110",
    year = "2022"
}

@ARTICLE{Gialamas:2024lyw,
       author = {{Gialamas}, Ioannis D. and {H{\"u}tsi}, Gert and {Kannike}, Kristjan and {Racioppi}, Antonio and {Raidal}, Martti and {Vasar}, Martin and {Veerm{\"a}e}, Hardi},
        title = "{Interpreting DESI 2024 BAO: late-time dynamical dark energy or a local effect?}",
      journal = {arXiv e-prints},
         year = 2024,
        month = jun,
          eid = {arXiv:2406.07533},
        pages = {arXiv:2406.07533},
          doi = {10.48550/arXiv.2406.07533},
archivePrefix = {arXiv},
       eprint = {2406.07533},
 primaryClass = {astro-ph.CO},
       adsurl = {https://ui.adsabs.harvard.edu/abs/2024arXiv240607533G}
}

@article{Colgain:2024xqj,
    author = "\'O Colg{\'a}in, Eoin and Dainotti, Maria Giovanna and Capozziello, Salvatore and Pourojaghi, Saeed and Sheikh-Jabbari, M. M. and Stojkovic, Dejan",
    title = "{Does DESI 2024 confirm {\ensuremath{\Lambda}}CDM?}",
    eprint = "2404.08633",
    archivePrefix = "arXiv",
    primaryClass = "astro-ph.CO",
    doi = "10.1016/j.jheap.2025.100428",
    journal = "JHEAp",
    volume = "49",
    pages = "100428",
    year = "2026"
}

@article{Favale:2024sdq,
    author = "Favale, Arianna and G\'omez-Valent, Adri\`a and Migliaccio, Marina",
    title = "{Quantification of 2D vs 3D BAO tension using SNIa as a redshift interpolator and test of the Etherington relation}",
    eprint = "2405.12142",
    archivePrefix = "arXiv",
    primaryClass = "astro-ph.CO",
    doi = "10.1016/j.physletb.2024.139027",
    journal = "Phys. Lett. B",
    volume = "858",
    pages = "139027",
    year = "2024"
}

@article{Huang:2024qno,
    author = "Huang, Zhiqi and others",
    title = "{Key drivers of the preference for dynamic dark energy}",
    eprint = "2405.03983",
    archivePrefix = "arXiv",
    primaryClass = "astro-ph.CO",
    reportNumber = "MEET-U-01",
    doi = "10.1103/PhysRevD.110.123512",
    journal = "Phys. Rev. D",
    volume = "110",
    number = "12",
    pages = "123512",
    year = "2024"
}

@article{Planck:2018vyg,
    author = "Aghanim, N. and others",
    collaboration = "Planck",
    title = "{Planck 2018 results. VI. Cosmological parameters}",
    eprint = "1807.06209",
    archivePrefix = "arXiv",
    primaryClass = "astro-ph.CO",
    doi = "10.1051/0004-6361/201833910",
    journal = "Astron. Astrophys.",
    volume = "641",
    pages = "A6",
    year = "2020",
    note = "[Erratum: Astron.Astrophys. 652, C4 (2021)]"
}

@article{Riess:2021jrx,
    author = "Riess, Adam G. and others",
    title = "{A Comprehensive Measurement of the Local Value of the Hubble Constant with 1 km s$^{−1}$ Mpc$^{−1}$ Uncertainty from the Hubble Space Telescope and the SH0ES Team}",
    eprint = "2112.04510",
    archivePrefix = "arXiv",
    primaryClass = "astro-ph.CO",
    doi = "10.3847/2041-8213/ac5c5b",
    journal = "Astrophys. J. Lett.",
    volume = "934",
    number = "1",
    pages = "L7",
    year = "2022"
}

@article{Freedman:2021ahq,
    author = "Freedman, Wendy L.",
    title = "{Measurements of the Hubble Constant: Tensions in Perspective}",
    eprint = "2106.15656",
    archivePrefix = "arXiv",
    primaryClass = "astro-ph.CO",
    doi = "10.3847/1538-4357/ac0e95",
    journal = "Astrophys. J.",
    volume = "919",
    number = "1",
    pages = "16",
    year = "2021"
}

@article{Pesce:2020xfe,
    author = "Pesce, D. W. and others",
    title = "{The Megamaser Cosmology Project. XIII. Combined Hubble constant constraints}",
    eprint = "2001.09213",
    archivePrefix = "arXiv",
    primaryClass = "astro-ph.CO",
    doi = "10.3847/2041-8213/ab75f0",
    journal = "Astrophys. J. Lett.",
    volume = "891",
    number = "1",
    pages = "L1",
    year = "2020"
}

@article{Blakeslee:2021rqi,
    author = "Blakeslee, John P. and Jensen, Joseph B. and Ma, Chung-Pei and Milne, Peter A. and Greene, Jenny E.",
    title = "{The Hubble Constant from Infrared Surface Brightness Fluctuation Distances}",
    eprint = "2101.02221",
    archivePrefix = "arXiv",
    primaryClass = "astro-ph.CO",
    doi = "10.3847/1538-4357/abe86a",
    journal = "Astrophys. J.",
    volume = "911",
    number = "1",
    pages = "65",
    year = "2021"
}

@article{Kourkchi:2020iyz,
    author = "Kourkchi, Ehsan and Tully, R. Brent and Anand, Gagandeep S. and Courtois, Helene M. and Dupuy, Alexandra and Neill, James D. and Rizzi, Luca and Seibert, Mark",
    title = "{Cosmicflows-4: The Calibration of Optical and Infrared Tully\textendash{}Fisher Relations}",
    eprint = "2004.14499",
    archivePrefix = "arXiv",
    primaryClass = "astro-ph.GA",
    doi = "10.3847/1538-4357/ab901c",
    journal = "Astrophys. J.",
    volume = "896",
    number = "1",
    pages = "3",
    year = "2020"
}

@article{Lee:2022cyh,
    author = "Lee, Bum-Hoon and Lee, Wonwoo and \'O Colg\'ain, Eoin and Sheikh-Jabbari, M. M. and Thakur, Somyadip",
    title = "{Is local H $_{0}$ at odds with dark energy EFT?}",
    eprint = "2202.03906",
    archivePrefix = "arXiv",
    primaryClass = "astro-ph.CO",
    doi = "10.1088/1475-7516/2022/04/004",
    journal = "JCAP",
    volume = "04",
    number = "04",
    pages = "004",
    year = "2022"
}

@article{Chevallier:2000qy,
    author = "Chevallier, Michel and Polarski, David",
    title = "{Accelerating universes with scaling dark matter}",
    eprint = "gr-qc/0009008",
    archivePrefix = "arXiv",
    doi = "10.1142/S0218271801000822",
    journal = "Int. J. Mod. Phys. D",
    volume = "10",
    pages = "213--224",
    year = "2001"
}

@article{Linder:2002et,
    author = "Linder, Eric V.",
    title = "{Exploring the expansion history of the universe}",
    eprint = "astro-ph/0208512",
    archivePrefix = "arXiv",
    doi = "10.1103/PhysRevLett.90.091301",
    journal = "Phys. Rev. Lett.",
    volume = "90",
    pages = "091301",
    year = "2003"
}

@article{DiValentino:2021izs,
    author = "Di Valentino, Eleonora and Mena, Olga and Pan, Supriya and Visinelli, Luca and Yang, Weiqiang and Melchiorri, Alessandro and Mota, David F. and Riess, Adam G. and Silk, Joseph",
    title = "{In the realm of the Hubble tension\textemdash{}a review of solutions}",
    eprint = "2103.01183",
    archivePrefix = "arXiv",
    primaryClass = "astro-ph.CO",
    reportNumber = "IPPP/20/108",
    doi = "10.1088/1361-6382/ac086d",
    journal = "Class. Quant. Grav.",
    volume = "38",
    number = "15",
    pages = "153001",
    year = "2021"
}

@article{Perivolaropoulos:2021jda,
    author = "Perivolaropoulos, Leandros and Skara, Foteini",
    title = "{Challenges for \ensuremath{\Lambda}CDM: An update}",
    eprint = "2105.05208",
    archivePrefix = "arXiv",
    primaryClass = "astro-ph.CO",
    doi = "10.1016/j.newar.2022.101659",
    journal = "New Astron. Rev.",
    volume = "95",
    pages = "101659",
    year = "2022"
}

@article{OColgain:2021pyh,
    author = "{\'O} Colg{\'a}in, Eoin and Sheikh-Jabbari, M. M.",
    title = "{Elucidating cosmological model dependence with $H_0$}",
    eprint = "2101.08565",
    archivePrefix = "arXiv",
    primaryClass = "astro-ph.CO",
    doi = "10.1140/epjc/s10052-021-09708-2",
    journal = "Eur. Phys. J. C",
    volume = "81",
    number = "10",
    pages = "892",
    year = "2021"
}

@article{Abdalla:2022yfr,
    author = "Abdalla, Elcio and others",
    title = "{Cosmology intertwined: A review of the particle physics, astrophysics, and cosmology associated with the cosmological tensions and anomalies}",
    eprint = "2203.06142",
    archivePrefix = "arXiv",
    primaryClass = "astro-ph.CO",
    reportNumber = "FERMILAB-CONF-22-192-SCD",
    doi = "10.1016/j.jheap.2022.04.002",
    journal = "JHEAp",
    volume = "34",
    pages = "49--211",
    year = "2022"
}

@ARTICLE{Colgain:2024ksa,
       author = {{\'O Colg{\'a}in}, Eoin and {Pourojaghi}, Saeed and {Sheikh-Jabbari}, M.~M.},
        title = "{Implications of DES 5YR SNe Dataset for $\Lambda$CDM}",
archivePrefix = {arXiv},
       eprint = {2406.06389},
 primaryClass = {astro-ph.CO},
journal={Eur. Phys. J \textbf{C}}, 
 volume = "85", 
doi = "10.1140/epjc/s10052-025-13995-4",
number ="286",
pages="286",
year="2025"
}

@article{Scolnic:2021amr,
    author = "Scolnic, Dan and others",
    title = "{The Pantheon+ Analysis: The Full Data Set and Light-curve Release}",
    eprint = "2112.03863",
    archivePrefix = "arXiv",
    primaryClass = "astro-ph.CO",
    doi = "10.3847/1538-4357/ac8b7a",
    journal = "Astrophys. J.",
    volume = "938",
    number = "2",
    pages = "113",
    year = "2022"
}

@ARTICLE{DESI:2024hhd,
       author = {{Adame}, A.~G. and others},
       collaboration="{DESI}",
        title = "{DESI 2024 VII: Cosmological Constraints from the Full-Shape Modeling of Clustering Measurements}",
      journal = {arXiv e-prints},
         year = 2024,
        month = nov,
          eid = {arXiv:2411.12022},
        pages = {arXiv:2411.12022},
          doi = {10.48550/arXiv.2411.12022},
archivePrefix = {arXiv},
       eprint = {2411.12022},
 primaryClass = {astro-ph.CO},
       adsurl = {https://ui.adsabs.harvard.edu/abs/2024arXiv241112022D}
}

@article{Gomez-Valent:2022hkb,
    author = "G{\'o}mez-Valent, Adri{\`a}",
    title = "{Fast test to assess the impact of marginalization in Monte~Carlo analyses and its application to cosmology}",
    eprint = "2203.16285",
    archivePrefix = "arXiv",
    primaryClass = "astro-ph.CO",
    doi = "10.1103/PhysRevD.106.063506",
    journal = "Phys. Rev. D",
    volume = "106",
    number = "6",
    pages = "063506",
    year = "2022"
}

@article{Colgain:2023bge,
    author = "\'O Colg{\'a}in, Eoin. and Pourojaghi, Saeed and Sheikh-Jabbari, M. M. and Sherwin, Darragh",
    title = "{A comparison of Bayesian and frequentist confidence intervals in the presence of a late Universe degeneracy}",
    eprint = "2307.16349",
    archivePrefix = "arXiv",
    primaryClass = "astro-ph.CO",
    doi = "10.1140/epjc/s10052-024-13727-0",
    journal = "Eur. Phys. J. C",
    volume = "85",
    number = "2",
    pages = "124",
    year = "2025"
}

@ARTICLE{Rubin:2023ovl,
       author = {{Rubin}, David and {Aldering}, Greg and {Betoule}, Marc and {Fruchter}, Andy and {Huang}, Xiaosheng and {Kim}, Alex G. and {Lidman}, Chris and {Linder}, Eric and {Perlmutter}, Saul and {Ruiz-Lapuente}, Pilar and {Suzuki}, Nao},
        title = "{Union Through UNITY: Cosmology with 2,000 SNe Using a Unified Bayesian Framework}",
      journal = {arXiv e-prints},
         year = 2023,
        month = nov,
          eid = {arXiv:2311.12098},
        pages = {arXiv:2311.12098},
          doi = {10.48550/arXiv.2311.12098},
archivePrefix = {arXiv},
       eprint = {2311.12098},
 primaryClass = {astro-ph.CO},
       adsurl = {https://ui.adsabs.harvard.edu/abs/2023arXiv231112098R}
}

@article{Afroz:2025iwo,
    author = "Afroz, Samsuzzaman and Mukherjee, Suvodip",
    title = "{Hint towards inconsistency between BAO and Supernovae Dataset: The Evidence of Redshift Evolving Dark Energy from DESI DR2 is Absent}",
    eprint = "2504.16868",
    archivePrefix = "arXiv",
    primaryClass = "astro-ph.CO",
    month = "4",
    year = "2025"
}

@article{Huang:2025som,
    author = "Huang, Lu and Cai, Rong-Gen and Wang, Shao-Jiang",
    title = "{The DESI DR1/DR2 evidence for dynamical dark energy is biased by low-redshift supernovae}",
    eprint = "2502.04212",
    archivePrefix = "arXiv",
    primaryClass = "astro-ph.CO",
    doi = "10.1007/s11433-025-2754-5",
    journal = "Sci. China Phys. Mech. Astron.",
    volume = "68",
    number = "10",
    pages = "100413",
    year = "2025"
}

@article{Holsclaw:2010nb,
    author = "Holsclaw, Tracy and Alam, Ujjaini and Sanso, Bruno and Lee, Herbert and Heitmann, Katrin and Habib, Salman and Higdon, David",
    title = "{Nonparametric Reconstruction of the Dark Energy Equation of State}",
    eprint = "1009.5443",
    archivePrefix = "arXiv",
    primaryClass = "astro-ph.CO",
    reportNumber = "LA-UR-09-05888",
    doi = "10.1103/PhysRevD.82.103502",
    journal = "Phys. Rev. D",
    volume = "82",
    pages = "103502",
    year = "2010"
}

@article{Holsclaw:2010sk,
    author = "Holsclaw, Tracy and Alam, Ujjaini and Sanso, Bruno and Lee, Herbert and Heitmann, Katrin and Habib, Salman and Higdon, David",
    title = "{Nonparametric Dark Energy Reconstruction from Supernova Data}",
    eprint = "1011.3079",
    archivePrefix = "arXiv",
    primaryClass = "astro-ph.CO",
    reportNumber = "LA-UR-09-07764",
    doi = "10.1103/PhysRevLett.105.241302",
    journal = "Phys. Rev. Lett.",
    volume = "105",
    pages = "241302",
    year = "2010"
}

@article{Alfano:2025fyq,
    author = "Alfano, Anna Chiara and Cafaro, Carlo and Capozziello, Salvatore and Luongo, Orlando and Muccino, Marco",
    title = "{Investigating the cosmic distance duality relation with gamma-ray bursts}",
    eprint = "2509.09247",
    archivePrefix = "arXiv",
    primaryClass = "astro-ph.CO",
    doi = "10.1016/j.jheap.2025.100444",
    journal = "JHEAp",
    volume = "49",
    pages = "100444",
    year = "2026"
}

@article{Kanodia:2025jqh,
    author = "Kanodia, Brijesh and Upadhyay, Ujjwal and Tiwari, Yashi",
    title = "{Revisiting Cosmic Distance Duality with Megamasers and DESI DR2: Model Independent Constraints on Early-Late Calibration}",
    eprint = "2507.11518",
    archivePrefix = "arXiv",
    primaryClass = "astro-ph.CO",
    month = "7",
    year = "2025"
}

@article{Zhang:2025qbs,
    author = "Zhang, Xuwei and Yang, Xiaofeng and Ren, Yunliang and Chen, Shuangnan and Shi, Yangjun and Cheng, Cheng and He, Xiaolong",
    title = "{Testing Cosmic Distance Duality Relation and Transparency with DESI DR2}",
    eprint = "2506.17926",
    archivePrefix = "arXiv",
    primaryClass = "astro-ph.CO",
    month = "6",
    year = "2025"
}

@article{Dhawan:2025mer,
    author = {Dhawan, S. and M{\"o}rtsell, E.},
    title = "{Implications for dark energy of cosmic transparency in light of DESI data}",
    eprint = "2506.22599",
    archivePrefix = "arXiv",
    primaryClass = "astro-ph.CO",
    month = "6",
    year = "2025"
}

@article{Li:2025htp,
    author = "Li, Tian-Nuo and Du, Guo-Hong and Wu, Peng-Ju and Qi, Jing-Zhao and Zhang, Jing-Fei and Zhang, Xin",
    title = "{Testing the cosmic distance duality relation with baryon acoustic oscillations and supernovae data}",
    eprint = "2507.13811",
    archivePrefix = "arXiv",
    primaryClass = "astro-ph.CO",
    month = "7",
    year = "2025"
}

@article{Zheng:2025cgq,
    author = "Zheng, Jie and Qiang, Da-Chun and You, Zhi-Qiang and Kumar, Darshan",
    title = "{Quantifying the Impact of 2D and 3D BAO Measurements on the Cosmic Distance Duality Relation with HII Galaxy observation}",
    eprint = "2507.17113",
    archivePrefix = "arXiv",
    primaryClass = "astro-ph.CO",
    month = "7",
    year = "2025"
}

@article{Avila:2025sjz,
    author = "Avila, Felipe and Oliveira, Fernanda and Franco, Camila and Lopes, Maria and Holanda, Rodrigo and Nunes, Rafael C. and Bernui, Armando",
    title = "{Probing the Cosmic Distance Duality Relation via Non-Parametric Reconstruction for High Redshifts}",
    eprint = "2509.07848",
    archivePrefix = "arXiv",
    primaryClass = "astro-ph.CO",
    doi = "10.3390/universe11090307",
    journal = "Universe",
    volume = "11",
    number = "9",
    pages = "307",
    year = "2025"
}

@article{Seikel:2012uu,
    author = "Seikel, Marina and Clarkson, Chris and Smith, Mathew",
    title = "{Reconstruction of dark energy and expansion dynamics using Gaussian processes}",
    eprint = "1204.2832",
    archivePrefix = "arXiv",
    primaryClass = "astro-ph.CO",
    doi = "10.1088/1475-7516/2012/06/036",
    journal = "JCAP",
    volume = "06",
    pages = "036",
    year = "2012"
}

@article{Shafieloo:2012ht,
    author = "Shafieloo, Arman and Kim, Alex G. and Linder, Eric V.",
    title = "{Gaussian Process Cosmography}",
    eprint = "1204.2272",
    archivePrefix = "arXiv",
    primaryClass = "astro-ph.CO",
    doi = "10.1103/PhysRevD.85.123530",
    journal = "Phys. Rev. D",
    volume = "85",
    pages = "123530",
    year = "2012"
}

@article{Wang:2025gus,
    author = "Wang, Qiumin and Cao, Shuo and Jiang, Jianyong and Zhang, Kaituo and Jiang, Xinyue and Liu, Tonghua and Mu, Chengsheng and Cheng, Dadian",
    title = "{New tests of cosmic distance duality relation with DESI 2024 BAO observations}",
    eprint = "2506.12759",
    archivePrefix = "arXiv",
    primaryClass = "astro-ph.CO",
    month = "6",
    year = "2025"
}

@article{Mukherjee:2025ytj,
    author = "Mukherjee, Purba and Sen, Anjan A.",
    title = "{New expansion rate anomalies at characteristic redshifts geometrically determined using DESI-DR2 BAO and DES-SN5YR observations}",
    eprint = "2505.19083",
    archivePrefix = "arXiv",
    primaryClass = "astro-ph.CO",
    doi = "10.1088/1361-6633/ae082c",
    journal = "Rept. Prog. Phys.",
    volume = "88",
    number = "9",
    pages = "098401",
    year = "2025"
}

@article{Teixeira:2025czm,
    author = "Teixeira, Elsa M. and Giar\`e, William and Hogg, Natalie B. and Montandon, Thomas and Poudou, Ad\`ele and Poulin, Vivian",
    title = "{Implications of distance duality violation for the $H_0$ tension and evolving dark energy}",
    eprint = "2504.10464",
    archivePrefix = "arXiv",
    primaryClass = "astro-ph.CO",
    month = "4",
    year = "2025"
}

@article{DESI:2024mwx,
    author = "Adame, A. G. and others",
    collaboration = "DESI",
    title = "{DESI 2024 VI: cosmological constraints from the measurements of baryon acoustic oscillations}",
    eprint = "2404.03002",
    archivePrefix = "arXiv",
    primaryClass = "astro-ph.CO",
    reportNumber = "FERMILAB-PUB-24-0154-PPD",
    doi = "10.1088/1475-7516/2025/02/021",
    journal = "JCAP",
    volume = "02",
    pages = "021",
    year = "2025"
}

@article{Wolf:2025acj,
    author = "Wolf, William J. and Ferreira, Pedro G. and Garc{\'\i}a-Garc{\'\i}a, Carlos",
    title = "{Cosmological constraints on Galileon dark energy with broken shift symmetry}",
    eprint = "2509.17586",
    archivePrefix = "arXiv",
    primaryClass = "astro-ph.CO",
    month = "9",
    year = "2025"
}

@article{Notari:2024zmi,
    author = "Notari, Alessio and Redi, Michele and Tesi, Andrea",
    title = "{BAO vs. SN evidence for evolving dark energy}",
    eprint = "2411.11685",
    archivePrefix = "arXiv",
    primaryClass = "astro-ph.CO",
    doi = "10.1088/1475-7516/2025/04/048",
    journal = "JCAP",
    volume = "04",
    pages = "048",
    year = "2025"
}

@article{Efstathiou:2024xcq,
    author = "Efstathiou, George",
    title = "{Evolving dark energy or supernovae systematics?}",
    eprint = "2408.07175",
    archivePrefix = "arXiv",
    primaryClass = "astro-ph.CO",
    doi = "10.1093/mnras/staf301",
    journal = "Mon. Not. Roy. Astron. Soc.",
    volume = "538",
    number = "2",
    pages = "875--882",
    year = "2025"
}

@inproceedings{Trotta:2017wnx,
    author = "Trotta, Roberto",
    title = "{Bayesian Methods in Cosmology}",
    eprint = "1701.01467",
    archivePrefix = "arXiv",
    primaryClass = "astro-ph.CO",
    month = "1",
    year = "2017"
}

@article{ACT:2025fju,
    author = "Louis, Thibaut and others",
    collaboration = "ACT",
    title = "{The Atacama Cosmology Telescope: DR6 Power Spectra, Likelihoods and $Λ$CDM Parameters}",
    eprint = "2503.14452",
    archivePrefix = "arXiv",
    primaryClass = "astro-ph.CO",
    reportNumber = "FERMILAB-PUB-25-0071-PPD",
    month = "3",
    year = "2025"
}

@article{Wilks:1938dza,
    author = "Wilks, S. S.",
    title = "{The Large-Sample Distribution of the Likelihood Ratio for Testing Composite Hypotheses}",
    doi = "10.1214/aoms/1177732360",
    journal = "Annals Math. Statist.",
    volume = "9",
    number = "1",
    pages = "60--62",
    year = "1938"
}

\end{document}